\documentclass{ar2e-changed}

\usepackage{graphicx}
\usepackage{url}
\usepackage{color}
\usepackage{amsmath,amssymb}
\usepackage[hyperindex,breaklinks]{hyperref}

\newcommand{\GeV}{\mathrm{GeV}}

\newcommand{\Mpl}{{m_{\mathrm{Pl}}}}
\def\fun#1#2{\lower3.6pt\vbox{\baselineskip0pt\lineskip.9pt
 \ialign{$\mathsurround=0pt#1\hfil##\hfil$\crcr#2\crcr\sim\crcr}}}
\def\lesssim{\mathrel{\mathpalette\fun <}}
\def\gtrsim{\mathrel{\mathpalette\fun >}}

\begin{document}

\input epsf.tex    
\input epsf.def   

\input psfig.sty

\jname{Ann.\ Rev.\ Nucl.\ Part.\ Sci.}
\jyear{2009}
\jvol{}
\ARinfo{}


\title{The Physics of Cosmic Acceleration}

\markboth{Robert R.~Caldwell and Marc Kamionkowski}{The Physics
of Cosmic Acceleration}

\author{Robert R.~Caldwell \affiliation{Department of Physics
\& Astronomy, Dartmouth College, Hanover, NH  03755}
Marc Kamionkowski
\affiliation{California Institute of Technology, Mail Code
130-33, Pasadena, CA 91125}}



\begin{abstract}
The discovery that the cosmic expansion is accelerating has been followed by an
intense theoretical and experimental response in physics and astronomy. The
discovery implies that our most basic notions about how gravity work
are violated on cosmological distance scales. One simple fix is the introduction of
a cosmological constant into the field equations for general relativity. However,
the extremely small value of the cosmological constant, relative to theoretical
expectations, has led theorists to explore a wide variety of alternative
explanations that involve the introduction of an exotic
negative-pressure fluid or a modification
of general relativity. Here we briefly review the evidence for
cosmic acceleration.  We then survey some of the theoretical
attempts to account for it, including the cosmological constant,
quintessence and its variants, mass-varying neutrinos, and
modifications of general relativity, such as scalar-tensor and
$f(R)$ theories and braneworld scenarios.  We discuss
experimental and observational tests that may allow us to
distinguish between some of the theoretical ideas that have been
put forward.
\end{abstract}


\maketitle

\section{Introduction}

The cosmic-acceleration puzzle is among the most viscerally
compelling problems in physics. Our deepest intuition about
gravity---that all objects should be attracted to each
other---just simply does not apply at cosmological distance
scales. Rather than slowing, as Newtonian gravity predicts, the
relative velocities of distant galaxies are increasing. The
implication is that gravity behaves far differently than
we had previously thought or that some mysterious fluid (``dark
energy'') with exotic gravitational properties fills the
Universe. Either way, there is new physics beyond the four
fundamental forces described by the Standard Model and general
relativity (GR). Cosmic acceleration thus motivates a
considerable fraction of current physical-cosmology research,
and it has become a major focus of particle- and string-theory
efforts.

There had long been hints, stemming primarily from the disparity
between the values $\Omega_m \simeq 0.1-0.3$ of the
nonrelativistic-mass density found by dynamical measurements and
the theoretical preference for a flat Universe,
$\Omega_{\mathrm{tot}}=1$, that there might be a cosmological
constant. However, direct measurements with distant supernovae
of a negative deceleration parameter provided the ``shot heard
'round the world'' \cite{Perlmutter:1998np,Riess:1998cb}.
The case for an accelerated expansion was dramatically bolstered
in 2000 with the CMB discovery of a flat Universe
\cite{deBernardis:2000gy}. A combination of
observations, based on galaxy surveys, the Lyman-alpha forest, baryon acoustic
oscillations, but primarily the CMB, now provide constraints to cosmological
parameters at a precision almost unimaginable a decade ago.  The
evidence for cosmic acceleration exists now at the
$\gtrsim50\,\sigma$ level \cite{Dunkley:2008ie}. It can no
longer be ignored.

The simplest solution involves no more than the addition of a
cosmological constant $\Lambda$ [with units of curvature, or
$({\mathrm{length}})^{-2}$] to Einstein's equation. But the
value required to explain cosmic acceleration is, in units where
$G=c=\hbar=1$, of order $10^{-120}$.  This is not a problem in
the classical world, but the quantum-theory expectation is that
the cosmological constant should be of order unity, or possibly
zero, should some symmetry or dynamical mechanism operate.
The gravitational effects of a cosmological
constant are equivalent to those of the virtual particles that
continually pop in and out of existence in quantum field theory.
Renormalization allows us to choose the zero point of this
virtual-particle energy density, but doing so implies a
cancellation of terms in the fundamental Lagrangian to one part
in $10^{120}$.

Dark-energy theories dodge this question.
The effects of a cosmological constant in Einstein's
equation can also be mimicked precisely by a homogeneous fluid of energy
density $\rho_\Lambda = \Lambda c^4/8\pi G$ and pressure
$p_\Lambda =-\rho_\Lambda$.  Dark-energy theories postulate that
the vacuum itself does not gravitate (by virtue of some
unspecificed symmetry or dynamical mechanism), but that the Universe is filled
with dark energy, an exotic negative-pressure fluid that
provides the impetus for cosmic
acceleration.  Alternative-gravity theories
investigated in this connection propose that an
accelerated expansion may simply be the vacuum solution of the
theory. The aim of the vast observational/experimental
dark-energy effort is to determine the physics of cosmic
acceleration.

While dark-energy and/or alternative-gravity theories do away
with the need for a cosmological constant, those that have been
developed so far require (as we will see below) the introduction
of unusually tiny parameters and/or finely-tuned initial
conditions.  They also introduce a new question, the ``coincidence
problem;'' i.e., why is the Universe transitioning
from deceleration to acceleration so recently?  None of the current
models answer this question fully, although some (e.g., the
tracker-field models discussed below) address it.

Theorists may debate the relative merits of various
cosmic-acceleration theories---cosmological constant, dark
energy, alternative gravity, anthropic arguments, etc.---but it
is ultimately up to experiment to decide which is correct.  The
most telling empirical quantity in this regard is the
(effective) dark-energy equation-of-state parameter $w_Q \equiv
p_Q/\rho_Q$, where $p_Q$ and $\rho_Q$ are the dark-energy
pressure and energy density, respectively.  The parameter $w_Q$
can be determined from the expansion history; i.e., by how the
cosmic acceleration changes with time.  If cosmic acceleration
is due to a cosmological constant, then $w_Q=-1$, and the future
expansion is de Sitter-like (i.e., exponentially expanding).  In
constrast, dark-energy and alternative-gravity theories predict
$w_Q\neq-1$.  Current constraints are $w_Q\simeq-1\pm0.1$.  The
precise value of $w_Q$ (and its evolution with time) depends on the particular
cosmic-acceleration theory.  There is no consensus on how far
from $-1$ it should be, but some classification of theoretical
predictions can be  provided.  Of course, cosmic-acceleration
theories require new physics, and this new physics may also be
probed experimentally in other ways, beyond just the expansion
history.

This review is intended primarily to survey some of the
theoretical explanations, involving both dark energy and
alternative gravity, for cosmic acceleration, and
secondarily to highlight the observational and experimental tests
that may be pursued to test the theories. We begin with some
background and a summary of the observational evidence 
for cosmic acceleration. We then review models that explain
cosmic acceleration by the introduction of a new exotic fluid
and those that work by modifying gravity. We close with a brief
review of some of the observational/experimental ways forward.
Refs.~\cite{Albrecht:2006um,Frieman:2008sn,Linder:2008pp}
complement this review by providing deeper analyses of observational
approaches to dark energy, while others
\cite{Peebles:2002gy,Padmanabhan:2002ji,Copeland:2006wr}
provide more details about recent dynamical models for dark energy.

\section{Background and Evidence}

\subsection{The Friedmann-Robertson-Walker cosmology}
We begin by reviewing the essentials of the standard
cosmological model.  We refer the reader to Ch.~13 in
Ref.~\cite{Peebles:1994xt} for more details.
\subsubsection{Kinematics}
An isotropic and homogeneous expanding Universe with spatial coordinates $x_i$ is
described by the Robertson-Walker metric,\footnote{Given the
observational evidence for negligible spatial curvature, we
assume throughout a flat Universe. This simplifies
considerably many of the equations. The effects of non-zero
curvature are discussed in Ref.~\cite{Caldwell:2004vi}.} $ds^2 =
-dt^2 +a^2(t) [dr^2 +r^2 (d\theta^2 +
\sin^2\theta d\phi^2)]$. The scale factor $a(t)$ is a function of time $t$, where
$a(t_0)=a_0$ at the present time $t_0$. Cosmologists use the redshift $z \equiv
(a_0/a) - 1$ as a proxy for the age or scale factor. The redshift can be measured
for distant sources; it is the fractional amount by which the wavelength of a photon
has been stretched by the expansion between the time the photon
is emitted and the time it is received.

The expansion rate $H\equiv \dot a/a$ is a function of time, with the value
$H_0\simeq70$~km~sec$^{-1}$~Mpc$^{-1}$ (the Hubble constant)
today, and where the dot denotes a derivative with respect to
$t$.  The deceleration
parameter is then $q \equiv -(\ddot a/a)/H^2 = (1+z) H'/H - 1$,
where the prime denotes a derivative with respect to $z$. The luminosity
distance of an object of luminosity $L$ at a redshift $z$ is defined to be
$d_L\equiv (L/4\pi F)^{1/2}$, where $F$ is the energy flux received from that
object. The luminosity distance is given (in a flat Universe) by
\begin{equation}
     d_L(z) = (1+z) c \int_0^z \,\frac{dz'}{ H(z')}.
\label{eqn:dL}
\end{equation}
Thus, measurement of the apparent brightness of sources of known luminosity
(``standard candles'') at a variety of redshifts $z$ can be used to determine or
constrain the expansion history.

The quantity $[H(z)(1+z)]^{-1} dz$ determines the time that evolves between
redshifts $z$ and $z+dz$; it thus also determines the physical
distance in this redshift interval and thus the physical volume
in a given redshift interval and angular aperture.
Likewise, $a(t)$ [which can be derived from $H(z)$] determines
the angular sizes of ``standard rods,'' objects of fixed
physical sizes. The angular-diameter distance is defined
to be $d_A(z)\equiv l_{\mathrm{prop}}/\theta$, where
$l_{\mathrm{prop}}$ is the proper size of an object and $\theta$
the angle it subtends on the sky.  It is related to $d_L$
through $d_A(z)=(1+z)^{-2} d_L(z)$.
As discussed below, measurements of the volume and luminosity
and angular-diameter distances can also be used to determine the
expansion history.

The integral expression for $d_L(z)$ can be Taylor expanded about $z=0$ to quadratic
order as $H_0 d_L(z) = cz[1+(1/2)(1-q_0)z+\cdots]$. The term linear in $z$ is the
well-known Hubble law. (Spatial curvature affects $d_L(z)$ only at cubic
or higher order \cite{Caldwell:2004vi}.) In 1998, two groups independently used
supernovae as standard candles to find better than $3\sigma$
evidence for a negative value for $q_0$
\cite{Perlmutter:1998np,Riess:1998cb}, the implications of which
will now be explained.

\subsubsection{Dynamics}
The Friedmann equation, 
\begin{equation}
     H^2 = \left( \frac{\dot a}{a} \right)^2 = \frac{8 \pi G}{3}\sum_i \rho_i,
\label{eqn:friedmann}
\end{equation}
is the general-relativistic equation of motion for $a(t)$ for a
flat Universe filled with fluids $i$ (e.g., nonrelativistic matter,
radiation, dark energy) of energy densities $\rho_i$. If
the fluids have pressures $p_i$, then the change $d(\rho a^3)$
in the total energy ($\rho=\sum_i \rho_i)$ per
comoving volume is equal to the work $-p d(a^3)$ (where
$p=\sum_i p_i$) done by the fluid. This relation
allows us to re-write the Friedmann equation as
\begin{equation}
     \frac{\ddot a}{a} = -\frac{4 \pi G}{3} \sum_i(\rho_i+3p_i).
\label{eqn:fried2}
\end{equation}
A nonrelativistic source has pressure $p=0$ implying $\ddot a <0$; i.e., the
relative velocities between any two galaxies should be
decreasing, in agreement with our Newtonian intuition.

If we define equation-of-state parameters $w_i\equiv p_i/\rho_i$
(e.g., $w_m=0$ for matter, $w_r=1/3$ for radiation), then the
second form, Eq.~(\ref{eqn:fried2}), of the Friedmann equation
can be written 
$q_0 = (1+3w_t)/2$, where $w_t\equiv p/\rho$ is the net
equation-of-state parameter.  Thus, if general relativity is
correct, the observations require that the Universe has
$w_t<-1/3$.  Thus, some ``dark energy,'' a negative-pressure
fluid, is postulated to account for cosmic acceleration.

\subsubsection{Expansion history}
Although the original supernova measurements determined only $q_0$,
future measurements will aim to determine the full functional dependence of $d_L(z)$
[or equivalently, $H(z)$] over the redshift range $0<z\lesssim$~few (the cubic
correction to $d_L(z)$ was first obtained observationally in
2004 \cite{Riess:2004nr}). If the
Universe consists today of nonrelativistic matter (baryons and dark matter; $w_m=0$)
with current energy density $\Omega_{m} \equiv \rho_{m}/\rho_c$ (in units of the
critical density $\rho_c =3H_0^2/8\pi G$) and some other exotic fluid with energy
density $\Omega_Q=1-\Omega_m$ and equation-of-state parameter $w_Q$, then 
$H(z) = H_0 [\Omega_m(1+z)^3 + (1-\Omega_m)(1+z)^{3(1+w_Q)}]^{1/2}$.
In a flat Universe, $\Omega_m+\Omega_Q=1$, the deceleration parameter is then
$q_0=(1+3w_Q \Omega_Q)/2$. The cosmological constant is equivalent to matter
with $w_Q=-1$; in this case, $q_0=(3/2)\Omega_m-1$.

\begin{figure}[!t]
\epsfscale1200		 
\centerline{\includegraphics[width=6in]{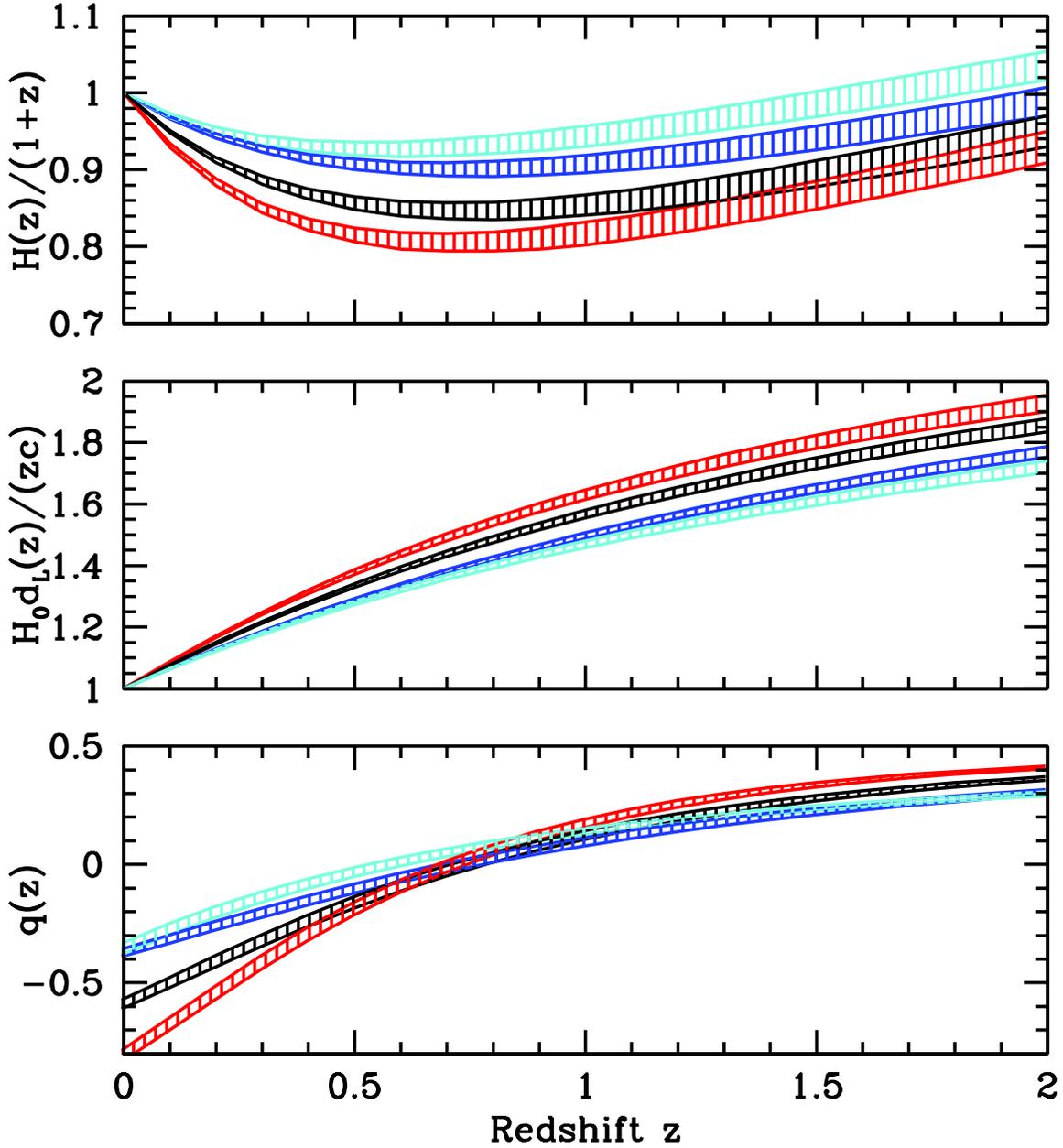}}
\caption{Examples of the expansion history $H(z)$, luminosity distance $d_L(z)$, and
   deceleration $q(z)$ are shown for several different dark-energy models. The red,
   black, and blue curves correspond to dark-energy models with
   equation-of-state parameter
   $w_Q=-1.2,\, -1,\, -0.8$, respectively. The cyan curve is for
   a DGP alternative-gravity model (see Sec.~\ref{sec:dgp}). All
   models have the same matter density and assume spatial
   flatness.    The thickness of the curves indicates the
   uncertainties that arise from the current uncertainty in the
   nonrelativistic-matter density $\Omega_m$.}
\label{figure:expansion}
\end{figure}

Note, however, that there is no reason to expect
$w_Q$ to be constant (unless $w_Q=-1$ precisely); it is just the simplest
parametrization of a time-varying dark-energy density. In much of the current
literature (including the Dark Energy Task Force Report \cite{Albrecht:2006um}), the
time evolution of $w_Q$ is parametrized as $w_Q=w_0+w_a(1-a/a_0)$. Most
generally, though, $w_Q(z)$ may be an arbitrary function of $z$; it is up to the
dark-energy theory (which we have not yet specified) to predict.
Fig.~\ref{figure:expansion} shows the expansion history
$H(z)$, luminosity distance $d_L(z)$, and deceleration
parameter $q(z)$ for four different models.  The first three
models are constant-$w_Q$ models with values of $w_Q=-1.2$,
$-1$, and $-0.8$. The fourth model is an
alternative-gravity model (DGP gravity), described in
Sec.~\ref{sec:dgp} below, with variable $w_Q$.  Measurement of
$w_Q(a)$ is the aim of observational efforts to probe the
physics of cosmic acceleration.

\subsubsection{Growth of Structure}

So far we have assumed that the Universe is perfectly
homogeneous, but this is only an approximation; the fractional
density perturbation $\delta_m(\vec x,t) \equiv [\rho_m(\vec
x,t)-\bar \rho_m]/\bar\rho_m$, where $\bar\rho_m$ is the mean density,
is not zero.  At sufficiently early times, or when smoothed on
sufficiently large scales, the fractional density perturbation
is $\delta_m \ll 1$.  In this linear regime, the density
perturbation satisfies an evolution equation,
\begin{equation}
     \ddot \delta_m + 2 H \dot \delta_m -(3/2) \Omega_m H^2
     \delta_m=0.
\label{eqn:growth}
\end{equation}
This equation has a
growing-mode solution (as a function of $z$) $\delta_m(z) \propto
D(z)$, and this evolution can be determined with
large-scale-structure measurements.  In the standard
cosmological model (i.e., $w_Q=-1$), the linear-theory growth
factor $D(z) \propto H(z)
(5\Omega_m/2) \int_z^\infty (1+z) [H(z)]^{-3}\, dz$.  This
expression is however invalid if $w_Q\neq-1$ \cite{Wang:1998gt},
and $D(z)$ will in general differ for different
$w_Q$.  Moreover, Eq.~(\ref{eqn:growth}) is derived assuming
that the dark energy remains perfectly homogeneous.  If dark
energy clusters, there may be a source for this equation (i.e.,
the right-hand side may be nonzero), in which case $D(z)$ may be
further affected.  Alternative theories of gravity invoked to
explain cosmic acceleration may predict a different $D(z)$, even
for the same expansion history.

\subsection{The Evidence}

\begin{figure}[!t]
\epsfscale1200		 
\centerline{\includegraphics[width=4in]{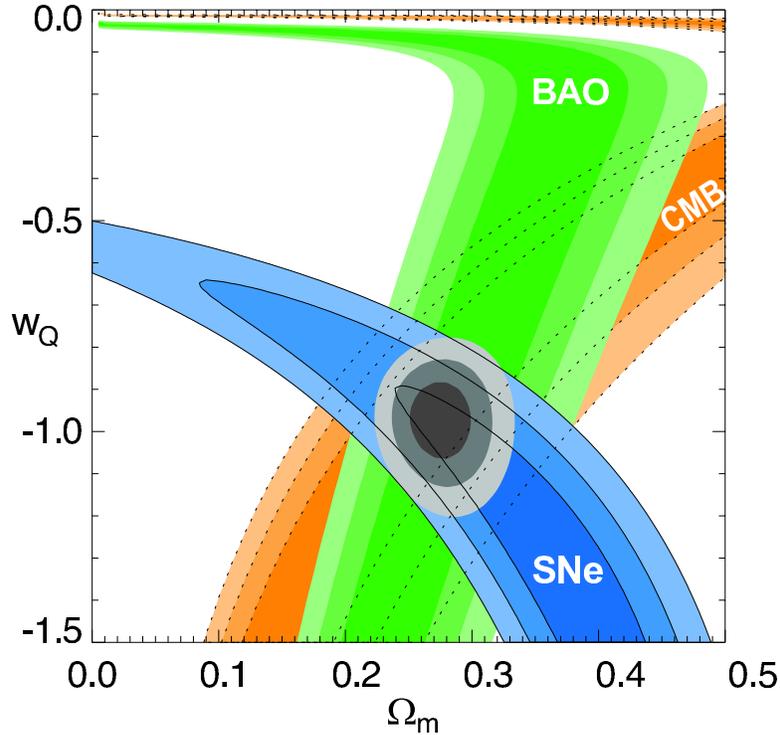}}
\caption{The $68.3\%$, $95.4\%$, and $99.7\%$ confidence-level
     contours for $w_Q$ and  $\Omega_m$ are shown, assuming a flat
     universe. The individual constraints from
     large-scale structure (using baryon acoustic oscillations), the cosmic
     microwave background, and the Union SN data set are shown, as well as the
     combined constraints. (From Ref.~\cite{Kowalski:2008ez}.)}
\label{figure:constraints}
\end{figure}

Evidence for accelerated expansion comes from the aforementioned direct
measurements of $d_L(z)$ using Type Ia supernovae, which now suggest $q_0\simeq
-0.7\pm0.1$ ($1\sigma$ errors) \cite{Kowalski:2008ez}.
However, the case for accelerated expansion is
dramatically bolstered by other observations.  Chief among these is the CMB
measurement of a flat Universe \cite{deBernardis:2000gy}, obtained by locating the
first acoustic peak in the CMB power spectrum \cite{Kamionkowski:1993aw}; this
implies a total density $\Omega_m+\Omega_Q\simeq1$ much greater than the matter
density $\Omega_m\simeq0.3$ indicated by dynamical measurements.  Current CMB
measurements alone are now sufficiently precise that they can determine a
dark-energy density $\Omega_Q= 0.742 \pm 0.030$ (for $w_Q=-1$
and a flat Universe) \cite{Dunkley:2008ie}, a
measurement that is made more precise with the addition of data from large-scale
structure, the Lyman-alpha forest, baryon acoustic oscillations, the cluster
abundance, and supernovae. In particular, supernova measurements
provide a constraint (again, assuming $w_Q=-1$) $q_0=(\Omega_m/2)-\Omega_Q
\simeq -0.7$ that
is nearly orthogonal to the CMB contour $\Omega_m+\Omega_Q \simeq 1$, and so
CMB and supernovae together provide tight limits in the
$\Omega_m$-$\Omega_Q$ plane.  The consistency of a
spatially-flat Universe with dark matter and a cosmological
constant with a wealth of precise data has led to the adoption
of a ``concordance model,'' our current standard
cosmological model.  Current values for the parameters of this
model are provided in Refs.~\cite{Dunkley:2008ie,Komatsu:2008hk}.

The concordance-model assumption $w_Q=-1$ can be tested
quantitatively with the data.  If
the Universe is flat, then the deceleration parameter is
$q_0 = (1/2)(1 + 3 w_Q \Omega_Q)$.  If $q_0<0$, then $w_Q < -(1/3)(1
-\Omega_m)^{-1}$, or $w_Q \lesssim -0.5$ for 
$\Omega_m\lesssim 0.3$.  The observed value $q_0\simeq -0.7$
requires an even more negative pressure, with $w_Q \simeq -1$.  The current
constraints to the $\Omega_m$-$w_Q$ parameter space, assuming a
constant $w_Q$ and flat Universe, are
shown in Figure~\ref{figure:constraints}.  Very
little is reliably known about the behavior of dark energy at $z\gtrsim 1$,
except that it does not appear to have played any significant role in cosmic
evolution at earlier times.  Other probes of the expansion
history will be discussed in Sec.~\ref{sec:observations}.

\section{Quintessence}

If the history of particle physics is any guide, then it seems reasonable
to speculate that the dark energy is due to a new field.
For cosmology, the simplest
field that can provide the missing energy between the matter density and
the critical density and drive cosmic acceleration is a scalar field. Such a field
in this role is sometimes referred to as ``quintessence'' to help distinguish it
from other fields or other forms of dark energy \cite{Caldwell:1997ii}.

\subsection{Basic Equations}  
\subsubsection{Background evolution}
The formal description of quintessence begins with the action,
\begin{equation}
     S = \int d^4x\, \sqrt{-g} \left( \frac{R}{16 \pi G} + {\cal L}_{SM} 
     + {\cal L}_Q \right),
\end{equation}
where $R$ is the Ricci scalar, $g$ the determinant of the metric.  Here,
the quintessence Lagrangian is ${\cal L}_Q =-\frac{1}{2}(\nabla_\mu Q)( 
\nabla^\mu Q) - V(Q)$, and ${\cal L}_{SM}$ is the Lagrangian for Standard Model
particles. The field obeys the Klein-Gordon equation, $\Box Q =
V_{,Q}$, where $\Box$ is the d'Alembertian and
$V_{,Q}\equiv \partial V/\partial Q$, and carries
stress-energy $T_{\mu\nu} = (\nabla_\mu Q) (\nabla_\nu Q) +
g_{\mu\nu}{\cal L}_Q$. (We
use metric signature $(-+++)$ and adopt the curvature conventions of
Ref.~\cite{Will:1993ns}.)

The spatially-homogeneous cosmic scalar is guided by the
equation of motion, $\ddot Q + 3 H \dot Q + V_{,Q}=0$,
with energy density and pressure,
\begin{equation}
     \rho_Q = \frac{1}{2}\dot Q^2 + V(Q), \qquad p_Q =
     \frac{1}{2}\dot Q^2 - V(Q).
\end{equation}
An equation-of-state parameter $w_Q<-1/3$ is obtained when $\dot
Q^2 < V$.  The mechanism for obtaining $\dot Q^2 \ll V$
is similar to the slow-roll mechanism in inflation (although not
precisely the same given that a fraction $\Omega_m\simeq0.25$ of
the current cosmological density is nonrelativistic matter).  We
illustrate here with the simple example of a potential $V(Q)
= (1/2)m^2 Q^2$.  In the absence of the Hubble-friction term
($3H\dot Q$) in the scalar-field equation of motion, the field will simply
oscillate in this quadratic potential.  However, if $m\ll H$,
then the Hubble friction overdamps the oscillator.  In this
case, $\ddot Q \ll H \dot Q, V_{,Q}$, and $3H\dot Q \simeq
-m^2Q$.  The field then moves little over a Hubble time, and
$\dot Q^2 \ll V$ is achieved.  More generally, quintessence
potentials are required to be very flat (have effective
masses $m_Q \equiv \sqrt{V_{,QQ}} \ll H$) to work.

\subsubsection{Expansion history and the quintessence potential}
More generally, a given quintessence potential
determines the expansion history and {\it vice versa}.  For
example, if quintessence has an
equation-of-state parameter $w(a)$ as a function of scale factor
$a$, then the energy density can be reconstructed as,
\begin{equation}
     \rho_Q(a) = \Omega_Q \rho_{c} 
     \exp{\left(3\int_a^{a_0} [1 + w(a)]\, d\ln a\right)} . 
\end{equation}
The potential and field evolution for this equation-of-state
parameter can then be reconstructed from,
\begin{eqnarray}
     V(a) &=&   \frac{1}{2}[1 - w(a)] \rho(a), \cr
     Q(a) &=&   \int d\tilde a {\sqrt{1 + w(\tilde a)}
     \over \tilde a H(\tilde a)} \sqrt{\rho(\tilde a)}.  
\end{eqnarray}
The equivalence $w(a) \leftrightarrow V(Q[a])$ is valid provided
$\dot Q\neq 0$. For most quintessence models, in which the field
evolves monotonically down a potential, this condition is satisfied.  

\subsubsection{Quintessence perturbations} 
If the quintessence field can vary in time, then it can most generally vary in
space. Linearized spatial fluctuations $\delta Q$ of the
quintessence field follow the evolution equation,
\begin{equation}
     \ddot{\delta Q} + 3 H \dot{\delta Q} + \left(V_{,QQ} 
     - \frac{1}{a^2}\nabla^2 \right) \delta Q 
     = \dot \delta_m \dot Q,
     \label{eqn:deltaQ}
\end{equation}
where $\nabla^2$ is the
spatial Laplace operator in comoving coordinates, and $\delta_m$
is the nonrelativistic-matter perturbation.  Quintessence
therefore responds to inhomogeneities in dark matter and baryons.
Furthermore, the source term depends on $\dot Q$, so
that the closer $w_Q$ is to $-1$, the weaker the driving term. The
nature of the response is determined by $m_Q$
or the quintessence Compton wavelength $\lambda_{Q} = m_Q^{-1}$.
In the case of constant $w_Q$, there is a simplification which 
can be written,
\begin{equation}
     V_{,QQ} =   - {3 \over 2}(1 - w_Q) \left[ \dot H - 
     \frac{3}{2}(1+w_Q)H^2 \right].
\end{equation}
For a slowly varying equation-of-state parameter, $V_{,QQ} \propto H^2$ and 
$\lambda_Q \sim H^{-1}$. From the above equations, this means that
fluctuations on scales smaller than the Hubble scale dissipate with sound speed
equal to the speed of light, since the coefficient of $\nabla^2/a^2$ in
Eq.~(\ref{eqn:deltaQ}) is unity, whence the field remains a smooth,
non-clustering component. Any initial fluctuations in the
quintessence field are damped out rapidly \cite{Dave:2002mn}.
In principle, perturbations to the quintessence field serve as a
source for matter perturbations---i.e., they show up as a non-zero
right-hand side to Eq.~(\ref{eqn:growth})---and thus affects the
linear-theory growth factor $D(z)$.  However, the damping of
small-scale quintessence perturbations implies that this is
generically a small effect.  On scales $\gtrsim H^{-1}$, the field is
gravitationally  unstable.  The growth of these long-wavelength
perturbations to quintessence may leave an imprint on the
large-angle CMB-anisotropy pattern, which we discuss in
Sec.~\ref{sec:observables}.

\begin{figure}[!t]
\epsfscale1200		 
\centerline{\includegraphics[width=6in]{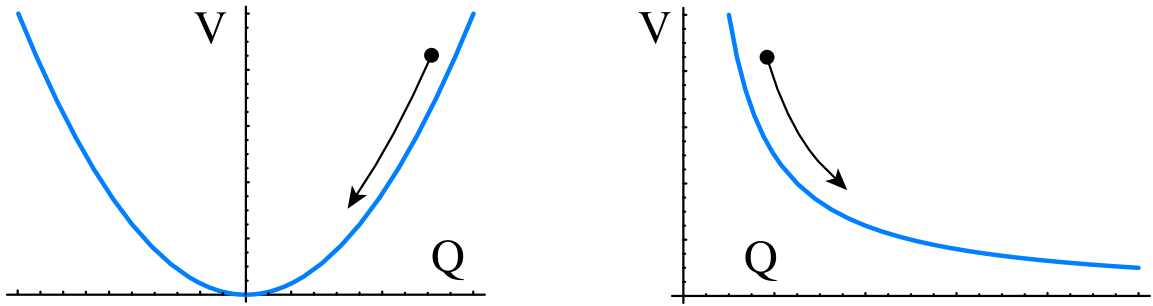}}
\caption{Two examples of potentials for the quintessence field. In the first,
representative of a conventional massive scalar or PNGB, the field is relaxing
towards the local minimum. In the second, representative of vacuumless potentials
such as the tracker, the field is evolving towards the global minimum.}
\label{figure:Q}
\end{figure}

\subsection{Representative Models}
Embedding scalar-field dark energy in a realistic extension of
the Standard Model faces a number of challenges. A viable
scenario generically requires an ultra-light
scalar ($m_Q  \lesssim H \sim 10^{-42}\,\GeV$), with
Planckian amplitude ($Q \sim 10^{19}\,\GeV$) and which remains
non-interacting with the Standard Model, and therefore
``dark'' \cite{Carroll:1998zi,Kolda:1998wq,Peccei:2000rz,Chung:2002xj}.  A
second challenge comes
from the coincidence problem---why is dark energy becoming
dominant today?
Ideally, the theory would have order-unity parameters at, say,
the Planck scale, and the dark-energy density today would be
insensitive to the field's initial conditions.  However,
in existing models, the parameters of the potential are
specially chosen
so that quintessence provides acceleration today.  Moreover, the scalar field
generically requires finely tuned initial conditions if the
field is to achieve the desired dynamics. Despite these
difficulties, many models of quintessence have been
proposed. Here we focus on just a few representative models.

\subsubsection{Cosmic Axion}  
A cosmic axion or pseudo-Nambu Goldstone boson (PNGB) is one
way to have a scalar of extremely low mass and keep it dark.
The first such models \cite{Frieman:1995pm} considered a
PNGB associated with a unification scale $f$ mediating a breakdown of a global
symmetry in a family of neutrinos at a scale $\mu \sim m_\nu^2/f$, thereby helping
to explain the very light mass of the quintessence field. Models have been proposed
employing string or M-theory moduli fields \cite{Choi:1999xn,Kim:2002tq}, too. The
resulting scalar potential, $V= \mu^4 (1 + \cos Q/f)$, is stable against loop
corrections (although not necessarily quantum-gravity effects
\cite{Kamionkowski:1992mf,Holman:1992us}), thereby protecting the mass $\mu$.
The shift symmetry, $Q \to Q+2 \pi f$, disables couplings to Standard Model
fields that would otherwise spoil the ``darkness''. A viable scenario requires $\mu
\simeq 0.002$~eV and $f\sim 10^{18}$~GeV
\cite{Coble:1996te,Dutta:2006cf,Abrahamse:2007te}. The cosmic evolution of the field
is as follows: the field has been frozen by Hubble friction through most of cosmic
history; as the Hubble friction relaxes, the field begins to
slowly relax towards its ground state, as illustrated in
Fig.~\ref{figure:Q}; in the future, the field will oscillate
at the bottom of
the potential, with its energy redshifting away like nonrelativistic matter. That
the mass scales $f$ and $\mu$ are derived from the energy scale  of other physics,
alleviates some need to explain the coincidence problem.  The
initial position of Q on the potential directly determines the
present-day properties of dark energy. However, the fine-tuning
problem is eased because the PNGB potential is periodic; the
range of starting values of Q that produces a viable scenario is
a non-negligible portion of the allowed range $Q\in [0,2\pi
f]$.

\subsubsection{Tracker Fields}
Condensation of hidden-sector quark--anti-quark pairs in a supersymmetric
version of QCD has been shown to give rise to a pion-like scalar field with an
effective potential $V = M^4 (Q/M_P)^{-n}$ with $n>0$
\cite{Affleck:1983mk,Binetruy:1998rz}. The index $n$ is
determined by the number of fermion families and colors, and $M$
is set by the cutoff scale. The cosmological dynamics of such a
scalar are quite novel
\cite{Peebles:1987ek,Ratra:1987rm,Zlatev:1998tr,Steinhardt:1999nw}: for a broad
range of initial conditions, the evolution of the field approaches, and then locks
onto, a universal track with negative equation-of-state parameter such that it inevitably
dominates at late times. When the scalar-field energy density is subdominant, its
equation-of-state parameter is $w_Q \approx (n w_B - 2)/(n + 2)$
where $w_B$ is the equation-of-state parameter of the dominant
or background component.  Hence, the field has been rolling
down the potential for most of its history, as illustrated in
Fig.~\ref{figure:Q}, but is now beginning to slow. As the
scalar field comes to dominate, its equation-of-state parameter
grows more negative, asymptoting to $w_Q\to -1$ in the
future. The universal track is uniquely
determined by the mass $M$ and index $n$, so that there is a one-to-one
relationship between $\Omega_Q$ and $w_Q$ as a function of
time. A viable model requires $0<n<1$ and $M\simeq
0.002$~eV. The
broad insensitivity of the late-time behavior to the initial conditions is
appealing---this model solves the above-mentioned fine-tuning
problem.  A further feature of the model is that it it goes
some way toward addressing the coincidence problem by allowing
the dark-energy density to track the matter/radiation density
over long periods of cosmological history.  
Still, there is no explanation as to why the
acceleration is happening now, as opposed to some later time.

\subsubsection{Exponential Potential}
A scalar field with an exponential potential, $V= M^4
e^{-\lambda Q/\Mpl}$, arises in a wide-variety of extensions of
Standard Model physics. In one particular case,
the scalar field is the dilaton, a pion-like condensate of
supersymmetric gaugino particles
\cite{Wetterich:1994bg,Binetruy:1998rz}. The dynamics of this
model are as follows: for $\lambda^2 > 3(1+w_B)$ the
scalar-field energy density tracks the background fluid with $w_Q=w_B$;
for $\lambda^2 < 2$ there are accelerating solutions
\cite{Copeland:1997et}. However, the scaling solutions do not
satisfactorily convert into dark energy at late times; a viable
model requires $\lambda^2<2$ and finely tuned initial conditions
for $Q$ and $\dot Q$.
Phenomenological variations on this model have been explored, for example, in
Ref.~\cite{Albrecht:1999rm}. These models feature a local
minimum in the exponentially decaying potential, where the field
can relax and produce potential-dominated accelerating expansion.

\subsubsection{Spintessence}
A scalar field with internal degrees of freedom has been considered as a dark-energy
candidate; one particular example is ``spintessence'' \cite{Boyle:2001du}, a complex
field $Q = R e^{i\Theta}$ spinning in a $U(1)$-symmetric potential $V=V(R)$. If the
spin frequency is high enough, $\dot\Theta \gg H$, then it is
rotation, rather than Hubble friction, that prevents the field
from rolling immediately to its minimum. The equation-of-state
parameter is $w \approx (R V'-V)/(R V'+V)$. Hence, a potential
with shape $R V' < V/2$
can provide $w_Q<-1/3$. However, the field is generically
unstable to the formation of Q-balls (non-topological solitons)
\cite{Boyle:2001du,Kasuya:2001pr}, rendering this solution to the dark-energy
problem unworkable. A gas of cold particles with an attractive interaction can also
yield negative pressure \cite{Nishiyama:2004ju}, but in the relativistic regime
required for cosmic acceleration, the theory winds up looking \cite{Fukuyama:2005jq}
like spintessence. A related idea is ``oscillescence,'' a single real scalar field
that oscillates in a confining potential $V(Q)\propto |Q|^n$.  This acts like a
fluid with $w_Q=(n-2)/(n+2)$ \cite{Turner:1983he} and thus gives
$w_Q<-1/3$ for $n<1$.  Again, though, this model is unstable to
small-scale perturbations \cite{Johnson:2008se}.

\subsubsection{K-essence} 
Dark-energy models with scalar degrees of freedom with
non-canonical kinetic terms in the Lagrangian display novel
dynamics. K-essence defines a class of models with a
Lagrangian ${\cal L}(Q,\,X)$ built from nonlinear functions of
$Q$ and $X \equiv -(1/2) (\nabla_\mu Q)( \nabla^\mu Q)$. The resulting
stress-energy tensor is $T_{\mu\nu} = {\cal
L}_{,X}(\nabla_\mu\phi)( \nabla_\nu \phi) + {\cal L}
g_{\mu\nu}$, so that the cosmic
pressure is simply $p={\cal L}$ and the energy density is $\rho
= 2 X p_{,X} - p$.
The motivation for these models is largely phenomenological, although in
string-inspired models the scalar is identified with the dilaton or other moduli
fields \cite{ArmendarizPicon:1999rj}. Purely kinetic k-essence, with ${\cal L}={\cal
L}(X)$, behaves as a barotropic fluid \cite{Scherrer:2004au}. A k-essence counterpart
to the potential-dominated tracker has ${\cal L}=f(\phi)(-X+X^2)$ with $f \propto
\phi^{-n}$. For $0< n < 2$, k-essence evolves with a constant
equation-of-state parameter
$w_Q=-1 +(n/2)(1+w_B)$ until it comes to dominate the Universe,
whereupon $w_Q\to -1$ \cite{Chiba:1999ka}. Models with multiple
attractor solutions, such that the field
scales with equation-of-state parameter $w_Q=1/3$ during the
radiation era, but then runs off to
a de Sitter-like solution after the onset of matter domination,
have been proposed
as possible solutions of the coincidence problem
\cite{ArmendarizPicon:2000dh}. However, there is another
aspect of k-essence that must be considered: the sound speed for the propagation of
high-frequency perturbations is $v^2 = p_{,X}/\rho_{,X}$. The canonical scalar field
has $v^2=1$.  The k-essence models that predict $v^2 <0$ are
eliminated because they are unstable to the growth of
fluctuations.  Density fluctuations in models with $0<v^2\ll 1$
can leave a strong imprint on the CMB and large-scale structure.
The apparent violation of causality in models with $v^2 > 1$,
including models that pass from scaling in the
radiation era to a present-day accelerating solution
\cite{Bonvin:2006vc}, suggests that additional analysis is
required to understand the phenomenology of these models
\cite{Babichev:2007dw}.
 
\subsubsection{Ghost Condensate} 
Dark-energy scalar-field theories built from higher-order
derivatives have also been studied. As an extension of
k-essence, these models are motivated by string field theory or
braneworld scenarios, too, and typically consist of a Lagrangian
that is a nonlinear function of $X$, $\Box Q$,
$(\nabla_\mu\nabla_\nu Q) ( \nabla^\mu \nabla^\nu Q)$, etc. In
certain cases, these higher-derivative terms can stabilize
theories with a leading-order kinetic term of the wrong sign
(hence, a ghost). One such case, a ghost condensate
\cite{ArkaniHamed:2003uy}, has equation-of-state parameter $w_Q=-1$ 
but carries fluctuations with a nonlinear dispersion relation
$\omega^2 \propto k^4$. This fluid contributes to the overall
inhomogeneous density field, yet the higher-derivative terms
mean that its fluctuations are sourced by higher derivatives of
the local gravitational fields. Generally, the additional
dynamics resulting from the higher-derivative terms allow novel
behavior, such as $w_Q \le -1$ with a stable, but vanishing
sound speed, $v^2\to 0$ \cite{Creminelli:2008wc}. Stable,
non-relativistic fluctuations contribute like a new species of
dark-matter inhomogeneties.

\begin{figure}[!t]
\epsfscale1200		 
\centerline{\includegraphics[width=6in]{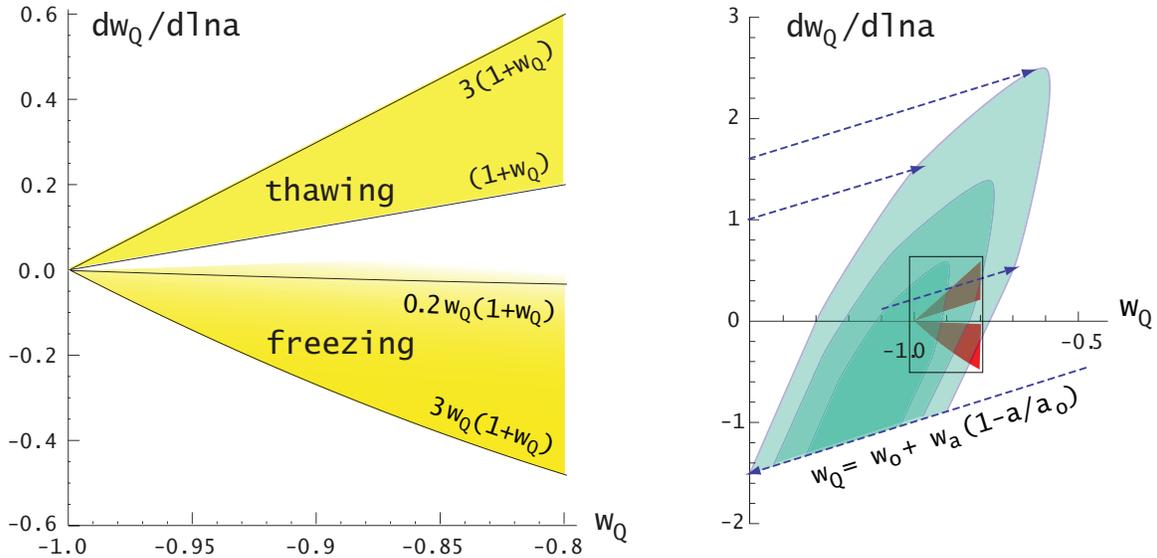}}
\caption{The $w_Q$ vs. $d w_Q/d\ln a$ parameter plane of
     dynamical dark-energy models is shown. On the left, the
     likely range of thawing and  freezing models is shown. On
     the right, the current  $68.3\%$, $95.4\%$, and $99.7\%$
     confidence-level constraints on the dark energy
     parameterization $w_Q = w_0 + w_a(1-a/a_0)$
     \cite{Kowalski:2008ez} have been converted into the
     present-day values of $w_Q,\, d w_Q/d\ln a$. The dashed lines
     show the direction of evolution of models located at
     particular points on the  $99.7\%$ confidence-level
     boundary.}
\label{figure:thawing}
\end{figure}

\subsection{Thawing and Freezing Models}  
The equation-of-state parameter for dynamical dark energy is
unlikely to be a constant. Using the cosmic axion and the tracker field as
guides, we may identify two classes of quintessence models, thawing and freezing.
Thawing models have a potential with a $V=0$ minimum accessible
within a finite range
of $Q$; the field starts high up the potential, frozen by Hubble
friction, with equation-of-state parameter $w_Q = -1$; as the Hubble constant decays, the field begins to thaw and roll
down towards $w_Q=0$. Freezing models are said to be vacuumless, as the minimum is not
accessible within finite range of $Q$, although there are no barriers; the field
rolls down the potential, but decelerates so that the
equation-of-state parameter evolves towards $w_Q\to -1$.  This
ignores models with nonzero local minima of the potential, but
these are equivalent to a cosmological constant with
massive-scalar-field excitations.
The trajectories of thawing and freezing models occupy rather
well-defined regions of the $w_Q$ vs. $d w_Q/d\ln a$ parameter plane
\cite{Caldwell:2005tm}, which are illustrated in Fig.~\ref{figure:thawing}. (Plenty
of models lie outside these regions \cite{Huterer:2006mv,Crittenden:2007yy},
although these tend to have metastable minima; e.g. a cosmological constant, or
non-canonical kinetic terms.)  These regions can be used as a guide for
assessing the sensitivity of methods to test for dynamical dark energy.

\subsection{Observables of the Models}
\label{sec:observables}
\begin{figure}[!t]
\epsfscale1200		 
\centerline{\includegraphics[width=4in]{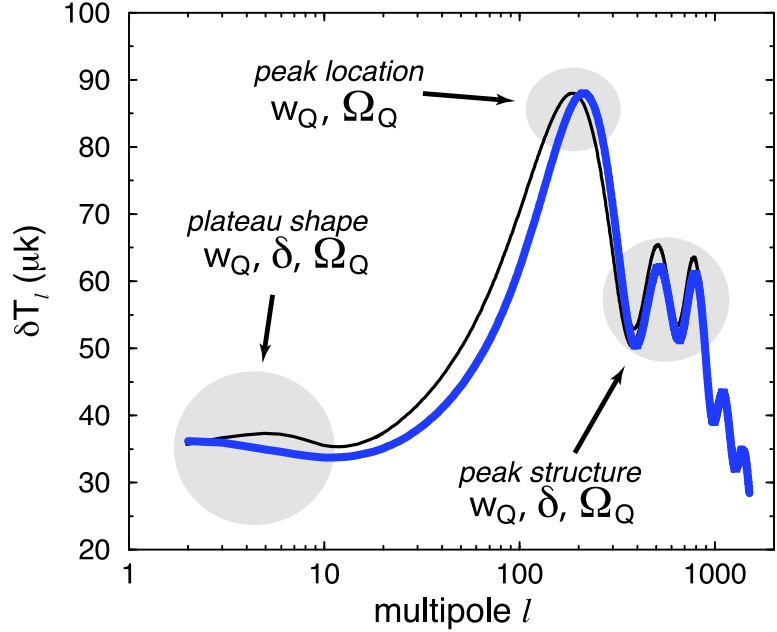}}
\caption{The effects of dynamical dark energy on the CMB
     temperature power spectrum are broadly
     illustrated. (1) Through the expansion history, the
     dark-energy density, noted in the figure by $\Omega_{Q}$
     and $w_{Q}$, influence the angular-diameter distance to last
     scattering, which sets the acoustic-peak multipole
     location. (2) Through the expansion history, dark energy
     influences the rate of growth of perturbations, affecting
     the CMB anisotropies created at late times on large angular
     scales. Fluctuations in the dark energy, noted by $\delta$,
     also contribute to the anisotropy pattern. (3) Dark energy
     can also influence the acoustic-peak structure if it has a
     non-negligible abundance at recombination \cite{Caldwell:2003vp}.}
\label{figure:qcmb}
\end{figure}

The dark-energy observables are the
energy density $\Omega_Q$, equation-of-state parameters $w_0$
and $w_a$, and the growth factor $D(z)$, which is determined by
the fluctuation sound speed $v$. 
The models described in this Section predict some time evolution
$a(t)$, and thus $w_0$ and $w_a$, and fluctuations that
propagate at a sound speed $v=1$ for quintessence, or more
generally $v\ge 0$ for k-essence. The dark-energy density and
equation-of-state parameter affect the
expansion history $H(z)$. The growth rate
of dark-matter and baryonic perturbations, as well as  the
gravitational potentials sampled by CMB photons, are sensitive
to the expansion history. Fluctuations in the dark energy,
dependent upon the dark-energy
density, equation-of-state parameter, and sound speed, can leave
an imprint on large-scale structure and the CMB.
The impact of these phenomena on the CMB power spectrum are illustrated
in Fig.~\ref{figure:qcmb}. 
A further discussion of observational
approaches will be given in Sec.~\ref{sec:observations}.

\section{Mass-Varying Neutrinos}

The coincidence
between the mass scale $m_\Lambda \equiv \Lambda^{1/4} \sim
10^{-3}$~eV of the cosmological constant and that of neutrino
masses motivates a solution that connects cosmic
acceleration to neutrino physics.  This idea was pursued in
Refs.~\cite{Fardon:2003eh,Peccei:2004sz} in
the idea of mass-varying neutrinos (MaVaNs).  Like quintessence,
the theory introduces a slowly-varying scalar field, dubbed the
acceleron, whose value determines the neutrino mass $m_\nu$.
The increased energy density associated with larger $m_\nu$
affects the acceleron dynamics in such a way that the
slow variation of the dark-energy density can be achieved with{\it out}
an extremely flat scalar-field potential.  

More quantitatively,
the energy density of the neutrino--dark-energy fluid is
$\rho_{\mathrm{dark}} = m_\nu n_\nu+\rho_{a}(m_\nu)$,
where $n_\nu$ is the neutrino number density and
$\rho_{a}$ the acceleron density, and the neutrinos
are assumed (here, for simplicity) to be nonrelativistic.  The
fields of the theory are designed so that the acceleron
relaxes to the value that minimizes
$\rho_{\mathrm{dark}}$, and the field value is thus fixed by the
condition $(\partial \rho_{\mathrm{dark}}/\partial m_\nu) =
n_\nu + (\partial \rho_{a}/\partial m_\nu)=0$.
Combining this with the energy-conservation equation, $\dot
\rho_{\mathrm{dark}} = -3 H (\rho_{\mathrm{dark}} +
p_{\mathrm{dark}})$, one finds that the dark-sector
equation-of-state parameter is
\begin{equation}
    w \equiv \frac{p_{\mathrm{dark}}}{\rho_{\mathrm{dark}}} =
    -1 + \frac{m_\nu n_\nu}{m_\nu n_\nu +\rho_{a}},
\label{eqn:MaVaNw}
\end{equation}
which gives $w\simeq-1$.

Specific implementations of
the theory may have testable consequences for
neutrino-oscillation experiments
\cite{Kaplan:2004dq}.
Unfortunately, however, the MaVaN idea suffers from a generic
instability \cite{Afshordi:2005ym,Brookfield:2005bz} to the growth of
perturbations that renders it unsuitable for explaining cosmic
acceleration.  The dark-energy density at any given
point is determined exclusively by the neutrino number density.
The gradient-energy density in this model is too small to
prevent the growth of spatial fluctuations.
Thus, the sound speed is $c_s^2=w<0$, giving rise
to a dynamical instability to the rapid growth of perturbations
to the MaVaN energy density.  A similar
instability arises generically in other models that similarly attempt
to couple dark matter and dark energy
\cite{Anderson:1997un,Bean:2007ny}.

\section{Phantom Energy}

The simplest dark-energy models (single-field models with
canonical kinetic terms) have $w_Q \ge -1$. However,
current data are consistent with $w_Q<-1$; for example, a recent
analysis finds $-1.14 < w_Q < -0.88$ (95\% CL)
\cite{Komatsu:2008hk}. It is thus interesting to ask, what if
the dark energy is phantom energy \cite{Caldwell:1999ew}? i.e.,
what if it has an equation-of-state parameter $w_Q<-1$?
Generally, dark energy with $w_Q<-1$ fits to 
observational data with a slightly lower energy density
$\Omega_Q$ than dark energy with $w_Q>-1$. There are important
differences, though, stemming from the fact that 
$w_Q<-1$ implies a violation of the null energy condition. This
means the energy density grows, rather than decays, with
time. In a phantom-dominated Universe, the scale factor
and expansion rate diverge in finite time, ripping apart everything---galaxies,
stars, atoms---before the Universe terminates in a ``big rip'' singularity
\cite{Caldwell:2003vq,McInnes:2001zw}. For example, assuming the equation-of-state
parameter maintains the constant value $w_Q=-1.1$, the end would
arrive in $\sim 100$~Gyrs. Spacetime diagrams classifying the
fate of the Universe with different types of dark energy were
presented in Ref.~\cite{Chiba:2005er}.

Theoretical models of phantom dark energy require exotic
physics, such as a scalar
field with negative kinetic energy or higher-derivative terms
\cite{Caldwell:1999ew,Carroll:2003st}. A quantum field with
negative kinetic energy is unstable; even if ``dark,'' then
gravitational interactions, unless cut off at a
sufficiently low energy scale
\cite{Carroll:2003st,Cline:2003gs}, can catalyze a
catastrophe. Curved-space quantum-field-theory models of phantom
energy are discussed in
Refs.~\cite{Parker:2001ws,Onemli:2004mb}. Quantum effects may
strengthen a big rip or a sudden singularity (a singularity in
which the scale factor remains finite but its derivatives
diverge) \cite{Barrow:2004xh} when the
spacetime-curvature radius shrinks to Planckian radius
\cite{Calderon:2004bi,Barrow:2008sn}. There
are other mechanisms that could masquerade as phantom energy,
such as novel photon \cite{Csaki:2004ha} or dark-matter
interactions \cite{Huey:2004qv}, as well as new gravitational phenomena \cite{Parker:2001ws}. Whereas a
canonical or k-essence scalar cannot cross the $w_Q=-1$ barrier
\cite{Vikman:2004dc,Hu:2004kh,Caldwell:2005ai}, such evolution
may be achieved in the presence of higher-derivative terms
\cite{Li:2005fm}.  

\section{Scalar-Tensor and f(R) Theories}

An alternative approach to cosmic acceleration is to change
gravity. 
With quintessence, we assume that general relativity is correct,
but that the Universe contains some exotic new substance that
drives cosmic acceleration; i.e., the left-hand side
($G_{\mu\nu}$) of Einstein's equation remains unaltered, but we
introduce a new source $T_{\mu\nu}$ for the right-hand side.
Here we alter general relativity; i.e.,
replace the left-hand side of Einstein's equation, or change
gravity even further.

\subsection{Scalar-tensor theories}

We begin by reviewing scalar-tensor theories, perhaps
the most widely studied class of alternative-gravity theories.
A wide array of experimental tests of such theories have been
investigated in detail \cite{Will:1993ns}.  Scalar-tensor
theories appear as low-energy limits of
string theories, and other alternative-gravity theories, such as
$f(R)$ theories (to be discussed below), can be recast as
scalar-tensor theories.  They can be understood heuristically as
models of gravity with a variable Newton's constant.

\subsubsection{The action and field equations}

In scalar-tensor theories, the Einstein-Hilbert action
$S_{\mathrm{EH}} =(16\pi G)^{-1} \int d^4x \sqrt{-g} R$ for
gravity is replaced by an action (see, e.g.,
Ref.~\cite{Carroll:2004st}),
\begin{equation}
    S=\int d^4x \sqrt{-g} \left[ b(\lambda) R -\frac{1}{2}
    h(\lambda) g^{\mu\nu} (\partial_\mu
    \lambda)(\partial_\nu\lambda) -U(\lambda) + {\cal
    L}_M(g_{\mu\nu},\psi_i) \right],
\label{eqn:STaction}
\end{equation}
where $\lambda(\vec x,t)$ is the eponymous scalar field; ${\cal
L}(g_{\mu\nu},\psi_i)$, the matter Lagrangian, is a function of
the metric and matter fields $\psi_i$; and $b(\lambda)$,
$h(\lambda)$, and $U(\lambda)$ are functions that determine the
form of the scalar-tensor theory.  The presence of a spatially
varying field $b(\lambda)$ multiplying the curvature $R$ in
Eq.~(\ref{eqn:STaction}) implies that scalar-tensor theories are
theories of gravity with a Newton's constant that depends on $b(\lambda)$.
The other terms in the action are then
kinetic- and potential-energy terms for the new field.  Although
Eq.~(\ref{eqn:STaction}) suggests that three functions
[$b(\lambda)$, $h(\lambda)$, and $U(\lambda)$] are required to
specify the theory, we can redefine $b$ to be the new field and
then derive new functions $U(b)$ and $h(b)$.

Variation of the action with
respect to the metric leads to the equation of motion (the
generalization of Einstein's equation),
\begin{equation}
    G_{\mu\nu} = b^{-1}(\lambda) \left[ \frac{1}{2}
    T_{\mu\nu}^{(M)} + \frac{1}{2} T_{\mu\nu}^{(\lambda)} +
    \nabla_\mu \nabla_\nu b - g_{\mu\nu} \Box b \right],
\label{eqn:STeinstein}
\end{equation}
where $G_{\mu\nu}$ is the Einstein tensor, $T_{\mu\nu}^{(M)}$
the stress tensor for matter, and
\begin{equation}
    T_{\mu\nu}^{(\lambda)} = h(\lambda) (\nabla_\mu\lambda)(
    \nabla_\nu \lambda) - g_{\mu\nu} \left[ \frac{1}{2}
    h(\lambda) g^{\rho\sigma} (\nabla_\rho
    \lambda)(\nabla_\sigma \lambda) + U(\lambda) \right],
\end{equation}
is the stress tensor for the scalar field.  There is also an
equation of motion,
\begin{equation}
    h \Box \lambda + \frac{1}{2} h' g^{\mu\nu} (\nabla_\mu
    \lambda) (\nabla_\nu\lambda) -U' + b' R=0,
\end{equation}
for the scalar field, where $'\equiv d/d\lambda$.

\subsubsection{Friedmann equations}

The equation of motion for the scale factor $a(t)$ in a
spatially-flat Robertson-Walker Universe is
\begin{equation}
    H^2  \equiv \left( \frac{\dot a}{a} \right)^2=
    \frac{\rho}{6b} + \frac{h \dot\lambda^2}{6b} - H
    \frac{\dot b}{b} + \frac{U}{6b},
\label{eqn:STFriedmann}
\end{equation}
with scalar-field equation of motion,
\begin{equation}
  \ddot \lambda + 3H \dot\lambda = 3 \frac{b'}{h} \left( \dot H
  +2 H^2\right) -\frac{h' \dot\lambda^2}{2h} -\frac{1}{2}
  \frac{U'}{h}.
\label{eqn:STeom}
\end{equation}
Several things are clear from these equations.  First,
there is considerable freedom in the choice of the functions
$b(\lambda$), $h(\lambda)$, and $U(\lambda)$, and so it is
difficult to make general statements about the validity of
scalar-tensor theories.  Secondly, specification of these
functions does not alone determine the phenomenology; the
initial conditions for the new scalar degree of freedom must
also be specified.

Although the detailed Friedmann and scalar-field equations are
different, there are explanations for cosmic acceleration in these
theories analogous to those in ordinary quintessence
theories.  For example, if $U(\lambda)$ is sufficiently shallow,
there may be solutions to the equations of motion in which
$\lambda$ is displaced from the minimum of $U(\lambda)$, and
rolls slowly.  In this case, the time derivatives in the
equations of motion will become negligible; the Friedmann
equation becomes approximately $H^2\simeq U/(6b)\simeq$constant;
and a roughly de Sitter expansion ensues.  Given the
additional terms in the Friedmann equation and scalar-field
equation of motion
that depend on derivatives of $b$ and $h$, the details may
differ, and a wider range of behaviors may be possible.
However, the form of the left-hand side of Eq.~(\ref{eqn:STeom})
implies that the rolling of the scalar field generically slows
with time; a general-relativistic cosmological behavior is
consequently an attractor in many scalar-tensor theories
\cite{Damour:1993id}.

\subsubsection{Brans-Dicke theory and Solar System tests}

The Brans-Dicke theory is defined by $b(\lambda)= \lambda/(16\pi G)$,
$h(\lambda)= \omega/(8\pi \lambda G)$, and $U(\lambda)=0$, where
the Brans-Dicke parameter $\omega$ is a constant.  Solution of
the field equations in the Solar System gives rise to a
parameterized-post-Newtonian (PPN) parameter
$\gamma=(\omega+1)/(\omega+2)$ for this theory.  This parameter
is measured in time-delay experiments in the Solar
System to be $\gamma=1 +(2.1\pm2.3)\times 10^{-5}$
\cite{Bertotti:2003rm}, leading to a bound
$\omega \gtrsim 5\times 10^4$.  The generalization of this
Solar System constraint to scalar-tensor theories with other
choices of $b(\lambda)$, $h(\lambda)$, and $U(\lambda)$ depends
on the specifics of those functions.  Roughly speaking, though,
the bound applies as long as the curvature at the minimum of
$U(\lambda)$ is sufficiently shallow so that the motion of
$\lambda$ within the Solar System is not restricted; this will
be quantified more precisely for $f(R)$ theories below.

\subsection{$f(R)$ Theories}

\subsubsection{The action and field equations}

One class of alternative-gravity theories that have received
considerable attention in recent years are $f(R)$ theories, in
which the Einstein-Hilbert action is replaced by an action
\cite{Capozziello:2003tk},
\begin{equation}
    S = \frac{1}{16\pi G} \int\, d^4x \sqrt{-g} f(R)
    +S_{\mathrm{matter}},
\label{eqn:fRaction}
\end{equation}
where $f(R)$ is a function whose form defines the theory.  Such
actions, which generalize the Einstein-Hilbert action, may arise
as low-energy limits of string theory.  Note that the $G$ in
this action is not necessarily the Newton's constant measured in
terrestrial experiments.

The field equations are obtained by varying the action with
respect to $g_{\mu\nu}$; the result is
\begin{equation}
    f'(R) R_{\mu\nu} -\frac{1}{2} f g_{\mu\nu} -\nabla_\mu
    \nabla_\nu f'(R) + \Box f'(R) g_{\mu\nu}
    =  8 \pi G T_{\mu\nu}.
\end{equation}
Taking the trace and setting $T_{\mu\nu}=0$, we find
a constant-curvature vacuum solution (i.e., a de Sitter
spacetime) with scalar curvature $R_0$, with $f'(R_0) R_0 =
2f(R_0)$.  

\subsubsection{Example: $1/R$ gravity}

In $1/R$ gravity \cite{Carroll:2003wy}, we choose $f(R) =
R-\mu^4/R$, with $\mu$ a
constant; this theory has a self-accelerating vacuum
solution with $R=12 H^2 =\sqrt{3} \mu^2$.  The field equation
for this theory is
\begin{equation}
  8\pi G T_{\mu\nu}= \left({1+\frac{\mu^4}{R^2}}\right)R_{\mu\nu} 
    - \frac{1}{2}\left({1-\frac{\mu^4}{R^2}}\right)Rg_{\mu\nu}
    +\mu^4 \left(g_{\mu \nu} \Box -
   \nabla_{(\mu}\nabla_{\nu)}\right) R^{-2}.
\label{fieldeq}
\end{equation}
Some intuition about the model can be obtained from the trace,
\begin{equation}
    \Box  \frac{\mu^4}{R^2} -  \frac{R}{3} +
    \frac{\mu^4}{R} = \frac{8 \pi G T}{3} , 
\label{eqn:trace}
\end{equation}
where $T=g^{\mu\nu}T_{\mu\nu}$.  For an effectively pressureless
source (e.g., the Sun), $T=-\rho$, where $\rho$ is the mass
density.  This should be compared with the general-relativistic
Einstein-equation trace, $R=8\pi G\rho$.  First of all, note
that (for constant $R$) the new equation is quadratic, rather
than linear, in $R$;
this suggests that there may be two different
constant-curvature solutions for the same $\rho$.  Given that $\mu^2
\sim H \ll G\rho$ in the Solar System, it is tempting to assume
an approximate ``GR-like'' solution $R\simeq 8\pi G\rho$.
However, this solution is violently unstable to
small-wavelength perturbations \cite{Dolgov:2003px} (and it
produces the wrong spacetime outside the Sun).  The other solution,
which has $R\simeq \mu^2$ everywhere throughout the Solar System
(and which is very different from $R\sim G\rho$), is stable.
However, this solution predicts a PPN parameter $\gamma=1/2$,
which disagrees with experimental constraints
\cite{Chiba:2003ir,Erickcek:2006vf}.

The other significant (and perhaps more important) difference
between $1/R$ gravity and general relativity arises from the term $\Box
(\mu^4/R^2)$ in Eq.~(\ref{eqn:trace}).  In general relativity, the trace
equation, $R=8\pi G \rho$, is a constraint equation that
determines $R$ uniquely.  However, in $1/R$ gravity, the scalar
curvature $R$ becomes a dynamical variable; i.e., there is a new
scalar degree of freedom.

The original $1/R$ gravity theory is just one example of an $f(R)$
theory.  Given the flexibility allowed in the choice of
$f(R)$, it is dangerous to draw general conclusions about $f(R)$
theories from $1/R$ gravity.  Still, there are several lessons:
(1) if the $f(R)$ theory is to explain cosmic
acceleration, there are likely to be mass parameters comparable
to $H$ in the theory.  (2) There is a scalar degree of freedom, dormant
in general relativity, that comes to life in $f(R)$ theories.  (3) Solar System
constraints to the theory may be severe. (4) There may be
more than one solution, for the same source, to the field
equations.  (5) One of the solutions may be unstable to
small-wavelength perturbations.

\subsubsection{The equivalence between $f(R)$ and scalar-tensor
theories}

In the general case, the physics of $f(R)$ theories can be
understood by noting that they are equivalent to scalar-tensor
theories \cite{Chiba:2003ir}.  Consider the following action for gravity with a
scalar field $\lambda$:
\begin{equation}
    S = \frac{1}{16 \pi G} \int d^4x \sqrt{-g} \left[ f(\lambda) +
    f'(\lambda)(R-\lambda) \right] + S_{\mathrm{matter}}.
\label{eqn:STequivalent}
\end{equation}
The $\lambda$ equation of motion gives $\lambda=R$ if
$f''(\lambda) \neq0$, demonstrating the equivalence with
Eq.~(\ref{eqn:fRaction}).  Eq.~(\ref{eqn:STequivalent}) is thus equivalent
to the scalar-tensor action, Eq.~(\ref{eqn:STaction}),
if we identify $b(\lambda)=f'(\lambda)$,
$U(\lambda) = -f(\lambda)+\lambda f'(\lambda)$, and
$h(\lambda)=0$.  In other words, $f(R)$ theories are equivalent
to scalar-tensor theories with vanishing kinetic term.
The absence of a kinetic term seems to suggest
that the scalar degree of freedom remains 
dormant, but if we change to an Einstein-frame metric
$g_{\mu\nu}^E = b'(\lambda) g_{\mu\nu}$ and canonical scalar field
$\varphi$ through $f'(\lambda) = \exp\left(\sqrt{16\pi G/3}
\varphi \right)$, then the Jordan-Brans-Dicke--frame (JBD-frame)
action [Eq.~(\ref{eqn:STequivalent})] becomes, in the Einstein
frame,
\begin{equation}
    S= \int d^4 x \sqrt{-g_E} \left[ \frac{1}{16\pi G} R_E
    -\frac{1}{2} g_E^{\mu\nu}
    (\partial_\mu\varphi)(\partial_\nu\varphi) - V(\varphi)
    \right],
\end{equation}
with
\begin{equation}
    V(\varphi) = \frac{ \lambda(\varphi) f'\left(\lambda(\varphi)
    \right) - f\left(\lambda(\varphi) \right)}{ 16\pi G
    [f'\left( \lambda(\varphi) \right)]^2}.
\label{eqn:Einsteinframe}
\end{equation}
In this frame, the propagating scalar degree of freedom is apparent.

In the Einstein frame, scalar-tensor theories look like
general relativity with a canonical scalar field.  The
difference, though, is that the Einstein-frame metric $g_E$ is
{\it not} the metric whose geodesics determine particle orbits;
it is the JBD-frame metric $g$.  Thus, scalar-tensor theories in
the Einstein frame resemble general relativity with an extra,
non-geodesic, force on the particle.  These may also be
generalized by ``chameleon theories'' \cite{Khoury:2003rn},
where the scalar coupling to matter may differ for
different matter fields.  Viewed in the JBD frame, 
scalar-tensor theories are those in which there is a new
propagating scalar degree of freedom, in addition to the usual
two propagating tensorial degrees of freedom.

\subsubsection{Friedmann equations}

The Friedmann equations for $f(R)$ theory are now obtained from
Eq.~(\ref{eqn:STFriedmann}),
\begin{equation}
    H^2 = \frac{1}{6} \frac{\rho}{f'(\lambda)} - H \frac{d}{dt}
    \ln f'(\lambda) + \frac{1}{6} \frac{\lambda
    f'(\lambda)-f(\lambda)}{f'(\lambda)},
\label{eqn:fRFriedmann}
\end{equation}
while the scalar-field equation of motion,
Eq.~(\ref{eqn:STeom}), now provides the constraint $\lambda =
12\, H^2 +6\, \dot H$.  The self-accelerating solution,
$H^2=\lambda/12$, can be obtained by setting $\rho=0$ and time
derivatives equal to zero in these equations.  We further see
that it is the scalar-field potential, $U(\lambda)= \lambda
f'(\lambda)-f(\lambda)$, in the last term of
Eq.~(\ref{eqn:fRFriedmann}) that is driving the accelerated
expansion.

\subsubsection{Solar System constraints}

The absence of a kinetic term for $\lambda$ implies a
Brans-Dicke parameter $\omega=0$ and thus a PPN parameter
$\gamma=1/2$ for $f(R)$ gravity, generalizing the
result for $1/R$ gravity.  However, as discussed above, this
Solar System constraint only applies if the function
$U(\lambda)$ in the scalar-tensor theory is sufficiently close
to flat that the scalar field can move freely in the Solar
System.  This is true if the following
four conditions are satisfied \cite{Chiba:2006jp}:
(i) $f(R)$ is analytic at $R=R_0$, where $R_0$ is the background
value of $R$; (ii) $f''(R_0) \neq 0$; and (iii)
$|f'(R_0)/f''(R_0)| r_{\mathrm{SS}}^2 \ll1$, where
where $r_{\mathrm{SS}} \sim$AU is the
distance scale over which the Solar System tests are carried
out; and (iv) $|f'(R_0)/f''(R_0)| \ll R_0(r_\odot/GM_\odot)$.
If these conditions are violated, then the linear-theory analysis that
concludes that $\gamma=1/2$ breaks down.  In this case, a fully
nonlinear analysis is required to determine $\gamma$.  

Theories can be constructed, by violating condition (iv) above,
that exhibit a ``chameleon mechanism,'' whereby the nonlinear
solution satisfies Solar System constraints
\cite{Starobinsky:2007hu,Faulkner:2006ub,Hu:2007nk}.  These
theories require the effective mass to be large in the Solar
System, and small in intergalactic space.  The GR-like
solution $R\simeq 8\pi G \rho$ inside the Sun matches onto the
GR-like solution with $\rho \simeq \rho_{{\mathrm{MW}}}$ (with
$\rho_{\mathrm{MW}}$ the mass density in the Milky Way) outside
the Sun, but within the Milky Way.  That solution then
transitions to the cosmological solution in intergalactic
space.  Functional forms for $f(R)$ that allow such behavior
require several small parameters.  Ref.~\cite{Amendola:2006we}
provides a classification of such models.
For example, an $f(R)$ that resembles a broken power
law, with $f(R) \propto R^n$ (with $n>0$) as $R\rightarrow 0$
and $f(R) \propto a + b/R^n$ (with $a$ and $b$ appropriately
chosen constants), may work \cite{Starobinsky:2007hu,Hu:2007nk}.
In these theories, the scalar-field dynamics
on cosmological scales become very ``stiff;'' i.e., the
phenomenology of these theories is almost indistinguishable from
those in which there is simply a cosmological constant
\cite{Appleby:2007vb}.

These models also imply a tail-wags-the-dog effect whereby a
change in the ambient density surrounding the Solar System, from
ISM densities to IGM densities, can change the results of PPN
tests by 5 orders of magnitude.  In some $f(R)$ theories,
particularly those with a chameleon mechanism, the usual $1/r^2$
force law of gravity is modified.  This seemingly trivial change
may have profound implications for almost every area of astrophysics,
from Solar System scales to the dynamics of galaxy clusters, few
of which have yet to be thought through carefully.

\subsubsection{Palatini formalism}
In the usual formulation (the ``metric formalism'') of general relativity, the
Einstein-Hilbert action is varied with respect to the metric
$g_{\mu\nu}$ to obtain Einstein's equations.  However, an
alternative approach, the ``Palatini formalism,'' is to vary the
action with respect to the connection $\Gamma^{\rho}_{\mu\nu}$
as well as the metric.
If applied to the Einstein-Hilbert action, this approach results
in the same gravitational field equations, and it also yields
the standard relation between the metric and the connection.
However, for a more general $f(R)$ action,
the Palatini formalism gives rise to a different theory.
Solutions to cosmic acceleration may also be obtained
with the Palatini formalism \cite{Vollick:2003aw}, possibly
without violating Solar System constraints. However,
the Christoffel symbol is now evaluated using a
different metric, $\tilde g_{\mu\nu} = f'(R) g_{\mu\nu}$,
whereas particle trajectories still follow geodesics of
$g_{\mu\nu}$.  Moreover, $R$ is obtained from the 
algebraic relation $R f'(R)-2 f(R) = 8 \pi G T$ between the Ricci scalar
and the trace of the stress-energy tensor. The gravitational
implications depend sensitively on the source stress tensor.
At the quantum level, these theories generally result in
new matter couplings that may have even more dire empirical
consequences \cite{Flanagan:2003rb} in the form of violations
of the equivalence principle \cite{Olmo:2005jd}.

\section{Braneworld Gravity and Related Ideas}

The alternative-gravity theories discussed above introduce a new
scalar degree of freedom.  Another possibility is to modify
gravity by changing the dimensionality of space.  We now discuss
such ``braneworld'' scenarios, as well as braneworld-inspired
ideas.  In braneworld scenarios, our $(3+1)$-d world is a
subspace of a higher-dimensional spacetime.  Unlike earlier
extra-dimensional models (e.g., Kaluza-Klein theories), Standard
Model fields may be restricted to lie on our brane, while
gravitational fields may propagate in the extra dimensions (the
``bulk'') as well.

\subsection{DGP Gravity}
\label{sec:dgp}

\subsubsection{The action}

DGP (for Dvali-Gabadadze-Porrati) gravity
\cite{Binetruy:1999ut,Dvali:2000hr} postulates a
$(4+1)$-dimensional Universe in which the
bulk of the five-dimensional spacetime is  Minkowski space with
an embedded $(3+1)$-dimensional brane (our Universe) on which
matter fields live.  The gravitational action is
\begin{equation}
    S_{(5)} = \int d^5x \sqrt{-g} \frac{R}{16\pi G^{(5)}} +
    \int d^4x \sqrt{-g^{(4)}} \left[
    \frac{R^{(4)}}{16\pi G} + {\cal L}_{SM} \right],
\label{eqn:DGPaction}
\end{equation}
where $G^{(5)}$ is the 5-d gravitational constant (note that its
dimensions are different than those of $G$), and
$g$ ($R$) the 5-d metric determinant (Ricci scalar), and
$g^{(4)}$ ($R^{(4)}$) the induced metric determinant (Ricci
scalar) on the brane.

\subsubsection{Heuristic picture}

Before proceeding with the cosmological solution, it is
instructive to consider DGP gravity in the weak-field limit.
If we take $g_{AB} = \eta_{AB} + h_{AB}$, with $|h_{AB}| \ll 1$,
then the linearized field equations tell us that
the 4-d metric components
$h_{\mu\nu}$, wherein resides the nonrelativistic potential, 
have Fourier ($p^\mu$) components,
\begin{equation}
    h_{\mu\nu}(p) = \frac{8\pi G}{p^2 + 2 (G/G^{(5)}) p}
    \left[ T_{\mu\nu}(p) - \frac{1}{3} \eta_{\mu\nu}
    T_\alpha^{\,\alpha}(p) \right],
\label{eqn:DGPlinear}
\end{equation}
for a stress-energy source $T_{\mu\nu}$ on the brane.  This
suggests a crossover distance, $r_0 = (1/2)(G^{(5)}/G)$.
For Fourier modes $p \gg r_0^{-1}$, $h_{\mu\nu}(p) \propto
p^{-2}$, implying the usual static gravitational potential $V(r) \propto
r^{-1}$ for $r \ll r_0$.  But for Fourier modes $p\ll r_0^{-1}$,
$h_{\mu\nu}(p) \propto p^{-1}$, implying $V(r) \propto r^{-2}$ at
larger distances; i.e., gravity is weaker at distances $r\gtrsim
r_0$.

\begin{figure}[!t]
\epsfscale1200		 
\centerline{\includegraphics[width=6in]{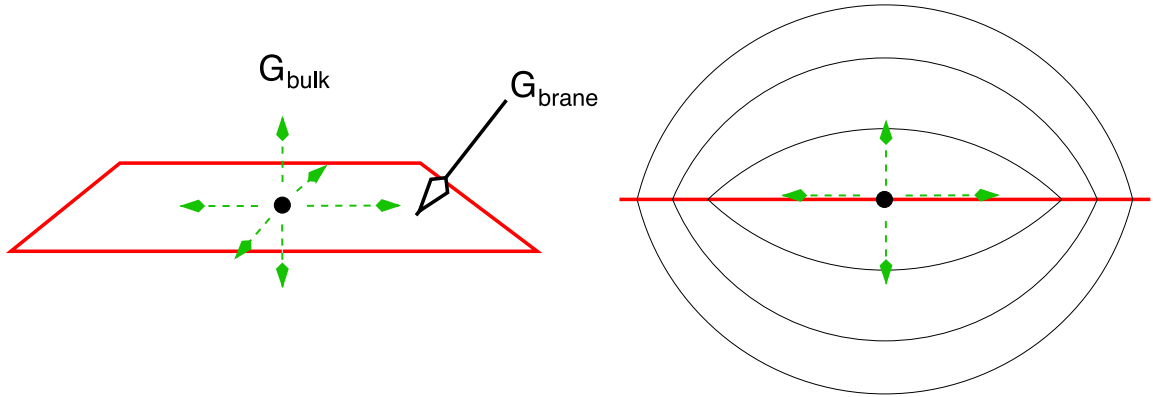}}
\caption{An illustration of the DGP mechanism.  Added to our
     $(3+1)$-d spacetime (the ``brane'') is an additional spatial
     dimension, the  ``bulk.''  The presence of stress-energy
     on the brane provides an energy cost for the propagation of
     gravitons with wavelengths $<r_0$ into the bulk,
     thus  making the gravitational force law $\propto r^{-2}$ at
      distances $r < r_0$ on the brane, but $\propto
     r^{-3}$ at distances $r< r_0$.  On the right are
     illustrated approximate equipotential curves for the
     gravitational field. (From Ref.~\cite{Lue:2005ya}.)}
\label{figure:dgp}
\end{figure}

The static gravitational potential in DGP gravity differs from
that in fundamental theories with small extra dimensions.
If there is an extra
dimension curled up into a size $R_5 \sim$mm, and the graviton
is free to propagate equally in our three spatial dimensions and
this extra small dimension, then the gravitational
force law steepens to $r^{-3}$ at distances $\lesssim$mm.  
In DGP gravity, however, the extra dimension is large, not
small, and there is an energy cost for the propagation of
gravitons with wavelengths $\lesssim r_0$ into the bulk.  At
$r\lesssim r_0$, the gravitons are thus confined to the brane,
and we have ordinary gravity; for $r\gtrsim r_0$, the gravitons
can escape into the bulk and the force law is that for a 5-d
spacetime, as shown in Fig.~\ref{figure:dgp}.

\subsubsection{Cosmological solution}

The action can be varied to obtain the field equations.  The
brane is then assumed to be filled with a homogeneous fluid
of pressure $p$ and energy density $\rho$
\cite{Binetruy:1999ut,Deffayet:2000uy}.  The cosmological
metric takes the form (assume a flat Universe for simplicity),
$ds^2 = N^2(t,\xi) dt^2 - A^2(t,\xi) d\vec x^2 - B^2(t,\xi) d\xi^2$,
where $\xi$ is the coordinate for the fifth dimension.  The
(generalized) Einstein equations yield equations of motion for
the metric variables $N(t,\xi)$, $A(t,\xi)$, and
$B(t,\xi)$.  The usual scale factor for our Universe is then
$a(t)=A(t,\xi=0)$, and it satisfies an equation of motion
(the DGP Friedmann equation)
\begin{equation}
    H^2 \pm \frac{H}{r_0} = \frac{8\pi G}{3} \rho.
\label{eqn:DGPFriedmann}
\end{equation}
There are two solutions for the expansion [cf., the discussion
of $f(R)$ models above].  If we take
the minus sign in Eq.~(\ref{eqn:DGPFriedmann}), then at early
times, when $H  \gg r_0^{-1}$, we recover the usual Friedmann
equation.  But when $H$ decreases, the new term kicks in, and $H
\rightarrow r_0^{-1}$ at late times; i.e., the Universe
asymptotes at late times to a de Sitter phase.  (The plus sign
in Eq.~(\ref{eqn:DGPFriedmann}) results in an eternally
decelerating Universe.)

\subsubsection{Solar System tests}

Unlike quintessence models, which retain general relativity, DGP gravity is an
alternative-gravity theory, and it makes predictions for
modified gravitational physics, beyond a modified expansion
rate, and in particular for a modified spacetime in the Solar System.
On the face of it, DGP gravity resembles at Solar System
distance scales a theory with a gravitational scalar degree of freedom.
This can be seen from the tensor structure, $T_{\mu\nu} - (1/3)
\eta_{\mu\nu} T_\mu^{\, \mu}$, that acts as the source for the
linearized gravitational field in Eq. (\ref{eqn:DGPlinear}).
This tensor structure resembles that in an $\omega=0$ scalar-tensor
theory and in massive gravity (the extra scalar being the
longitudinal mode of the graviton), but differs from the
structure $T_{\mu\nu}-(1/2) \eta_{\mu\nu} T_\mu^{\, \mu}$ in
general relativity.  The extra scalar degree of freedom in DGP gravity may be
understood as a fluctuation in the brane surface.  The difference
means that a relativistic particle (e.g., a photon) is affected
differently by the same source, leading to a PPN parameter
$\gamma=1/2$, again in disagreement with measurements.  This
is a DGP equivalent of the van
Dam-Veltman-Zakharov discontinuity
\cite{vanDam:1970vg,Zakharov:1970cc} that appears in massive gravity.

However, Eq.~(\ref{eqn:DGPlinear}) provides only the $(3+1)$-d
components of the field.  The approximations that lead to this
linearized equation involve a highly nonlinear metric
perturbation in the bulk, even when the source is weak, calling
the derivation of Eq.~(\ref{eqn:DGPlinear}) into question.  A
proper treatment involves a perturbative expansion not only in
$h_{\mu\nu}$, but also in $r/r_\ast$, where $r_\ast = (r_g
r_0^2)^{1/3}$, and $r_g =2 G M/c^2$ is the Schwarzchild radius
\cite{Gruzinov:2001hp}.  The
field equations for the spherically symmetric spacetime can then
be solved perturbatively in three different 
distance regimes, with the following results:  (1) the spacetime
looks like that in general relativity, with fractional
corrections $O\left( (r/r_\ast)^{3/2} \right)$, at small distances, $r\ll
r_\ast$; (2) it looks like that in an $\omega=0$ scalar-tensor
theory (i.e., the static potential is still $\propto r^{-1}$,
but light deflection is described by
$\gamma=1/2$) at distances $r_\ast \ll r \ll r_0$;
and (3) it then falls off more steeply, as $r^{-2}$, at distances $r
\gg r_0$.

For example, $r_\ast \simeq 150$ pc for the spacetime around the
Sun.  Thus, Solar System tests of gravity occur deep within the
GR-like regime, and DGP gravity is thus consistent with these
tests.  Still, the spacetime is not {\it precisely}
Schwarzchild; there are corrections $O\left((r/r_\ast)^{3/2} \right)$.  These
corrections may be tested by future experiments \cite{Lue:2002sw}, although the
$r^{-3/2}$ dependence of the correction means that the theory
can{\it not} be parametrized with the usual PPN formalism.  

Light-deflection experiments in the Solar System are unlikely to
be constraining, as the fractional correction to the
general-relativistic value
for the deflection angle will be $\sim (r/r_\ast)^{3/2}\sim
10^{-11}$, while the smallest value probed is $\sim10^{-4}$.
However, measurements of perihelion advances may be more promising.
DGP gravity leads to a correction, $\sim 5\,
\mu$as/year, to the perihelion advance of a planetary orbit
\cite{Lue:2002sw}.
Unlike the general-relativistic perihelion-advance rate, which decreases for
larger-$r$ orbits, the DGP correction is $r$ independent and can
thus be distinguished from a general-relativistic correction (or from those that
occur in the usual PPN expansion).  Moreover, Solar System tests
at large distances may be equally, or more, effective at testing
DGP gravity as those at short distances.  Thus, improved lunar
laser ranging experiments may be sensitive to DGP gravity
\cite{Dvali:2002vf}, as might BepiColombo and MESSENGER,
European Space Agency and NASA satellites, respectively, to
Mercury.  However, probes of the outer Solar System, like
Cassini, might also probe DGP gravity.

\subsubsection{Expansion history}

By re-arranging Eq.~(\ref{eqn:DGPFriedmann}), the expansion
history can be written, $H(z) = (H_0/2) \left[ 1-\Omega_m + \sqrt{
(1-\Omega_m)^2    + 4 \Omega_m (1+z)^3} \right]$.
At $z\gg1$, this approaches the standard form, $H(z) \simeq
\sqrt{\Omega_m}H_0 (1+z)^{3/2}$, and $w_Q\to-1$ in the distance
future, $z\to-1$.
The deceleration parameter for this model is $q_0 = 3\Omega_m
(1+\Omega_m)^{-1}-1$, and thus there is a relation between
$q_0$ and $\Omega_m$.  A value $\Omega_m=0.274$ implies
$q_0=-0.355$, which is only marginally consistent with current
data.  A better fit to observations can be obtained by
adding a cosmological constant or curvature
\cite{Deffayet:2002sp}, or in models based on other
manifestations of braneworlds.  

Fig.~\ref{figure:expansion} shows the expansion history,
luminosity distance, and deceleration for the DGP
model.  If the expansion history can be measured with sufficient
precision to distinguish this functional form from,  e.g., a
constant-$w_Q$ model, then this may provide an avenue toward
testing the model.

\subsubsection{Growth of structure}
The distance scales relevant for large-scale structure generally
occur at $r\gtrsim r_\ast$, where the behavior of
DGP gravity differs from that of general relativity.  The growth
of linear density
perturbations can be described in DGP gravity in terms of an
effective Newton's constant, $G_{\mathrm{eff}} =  G( 1+
3/\beta)$ \cite{Lue:2004rj}, with 
$\beta = 1 - 2 r_0 H \left[ 1 +\dot H/(3 H^2) \right]$.  The
effects of this altered gravitational constant can be taken into account
approximately by changing the last term in
Eq.~(\ref{eqn:growth}); the factor $\Omega_m H^2$ that
appears there arises from the Friedmann equation $\Omega_m H^2 =
8\pi G\rho/3$.  The change in the linear-theory growth factor
$D(z)$ can be appreciable in these models; it is a
$\sim 30\%$ correction at $z=0$.
This constrasts dramatically with quintessence models, which
do not generally affect $D(z)$ significantly.

\subsection{Related ideas}

There have been other attempts to modify gravity to account for
cosmic acceleration that are inspired by DGP gravity or the
massive-gravity theories it resembles.

\subsubsection{Degravitation}
The idea of degravitation
\cite{ArkaniHamed:2002fu,Dvali:2007kt} is to replace Einstein's
equation, $G_{\mu\nu} = 8\pi G T_{\mu\nu}$, by $\left[ 1 + F(L^2
\Box) \right]G_{\mu\nu} = 8 \pi G T_{\mu\nu}$,
where $F(x)$ is a monotonically decreasing filter function with
the limits $F(x)  \rightarrow  0$ for $x \rightarrow
\infty$, and $F(x) \gg 1$ for $x \rightarrow0$.
Here, $L$ is a distance scale (presumably $\sim H_0^{-1}$) at
which the force of gravity weakens.  Thus, Newton's constant
acts as a high-pass filter; 
long-wavelength modes of the stress-energy tensor do not source
the gravitational field.  An analogous modification of
electrodynamics is precisely equivalent to electrodynamics with
a massive photon.  Likewise, the structure of degravitation
shares some similarities with massive gravity, although the
mapping is not precise.  

\subsubsection{The fat graviton}  The idea here
\cite{Sundrum:2003jq} is to postulate that virtual gravitons with
invariant masses $\gtrsim$meV just simply do not propagate; the
cosmological constant due to zero-point fluctuations conveyed by
gravity is then that
observed.  Such models can be constrained by considering
cosmological gravitational-lensing systems
\cite{Caldwell:2006gu}, as the angular deflection of photons in
such systems implies momentum transfers (presumably carried by
virtual gravitons) larger than this energy scale.

\subsubsection{Modified Friedmann equations}

Braneworld scenarios that generalize the DGP theory
by allowing for a wider range of dynamics in the bulk can
produce an effective expansion law $H^2 \propto \rho^n$ on the
brane \cite{Sahni:2002dx}, and this has motivated phenomenological
models of dark energy.  One such example is the ``cardassian''
model whereby $H^2 =  (8 \pi G\rho/3) + B \rho^n$
\cite{Freese:2002sq}; supernova and CMB distances suggest $n
\lesssim 0.4$. An alternative parameterization of the effects of
extra dimensions proposes $H^2 + (1-\Omega_M)H_0^2
(H/H_0)^\alpha =  8 \pi G \rho/3$
\cite{Dvali:2003rk}. During the matter era, the
equation-of-state parameter of the inferred dark energy is
$w_{\mathrm{eff}} = -1 + \alpha/2$
until $z\sim 1$, and asymptotes to $w_{\mathrm{eff}}\to -1$ in the
future. Rough arguments suggest that $\alpha \lesssim 1$ is
necessary for consistency with observations.

\subsubsection{A phenomenological approach}

In Refs.~\cite{Bertschinger:2006aw,Caldwell:2007cw,Hu:2007pj} the
existence of a new gravitational theory is posited that changes
the amount of spacetime curvature produced per unit mass. The
Friedmann equation is modified so that the matter-dominated
expansion becomes progressively more de Sitter-like, mimicking
the evolution under dynamical dark energy with equation-of-state
parameter $w_Q \simeq -1$. Metric perturbations likewise respond
differently to inhomogeneities in the matter and radiation,
leading to a characteristic ``gravitational slip'' whereby the
potential $\psi$ appearing in the geodesic equation, $\ddot {\vec
x} = -\vec\nabla\psi$ differs from the potential $\phi$ in the
Poisson equation, $\nabla^2\phi = 4 \pi G \delta\rho$.
Scalar-tensor and $f(R)$ theories, braneworld scenarios, and DGP
gravity, as well as massive gravity, all predict $\phi \ne \psi$
in the presence of non-relativistic matter, in contrast to
general relativity. This suggests a parameterized post-Friedmann
description of modified gravity, whereby a new parameter,
$\varpi \equiv \psi/\phi - 1$ characterizes the degree of
departure from general relativity, in analogy to the
post-Newtonian parameter $\gamma$. The imposed time- and
scale-dependence of $\varpi$, along with two further
assumptions---conservation of the radiation and matter
stress-energy tensor, and the absence of new gravitational
effects mimicking a  ``dark fluid'' momentum flux or velocity
relative to the cosmic rest frame---are sufficient to complete
the description of linearized metric perturbations. A
$\varpi \neq 0$ affects the rate of growth of perturbations, the
integrated Sachs-Wolfe effect, and the
weak-gravitational-lensing deflection angle. Hence, observations
of the CMB and
large-scale structure may be used to test for the consistency of
general relativity on cosmological scales.

\subsection{Comments}

There are a number of theoretical questions that must be
addressed if braneworld scenarios are to explain cosmic
acceleration.  The simplest DGP model is only marginally
consistent with the observed cosmic
acceleration; either some new exotic fluid or a more complicated
implementation of the braneworld must be introduced to improve
the agreement with data.
Braneworld scenarios introduce new small parameters, and they do
not solve the coincidence problem.  Moreover, 
it has been questioned whether the local perturbative solutions
for the spherically symmetric DGP spacetime can be sewn together
into a single global solution \cite{Damour:2002gp}. There are
also questions about the stability of the self-accelerating
phase to the growth of small-scale fluctuations
\cite{Luty:2003vm,Gregory:2008bf}.  Still, braneworld scenarios
and related ideas are worth further theoretical attention as
they connect cosmology to novel ideas from string and
supergravity theories and
provide a range of phenomenological consequences, beyond simply
the alteration of the expansion rate for which they were
introduced to explain.

\section{The Landscape Scenario}
\label{sec:landscape}

We have concentrated on theories of cosmic
acceleration based on the introduction of new fields or
modifications of gravity, both intended as alternatives to the
simple postulate of a cosmological constant.  But
cosmic acceleration may just be due to
a cosmological constant.  If so, then the physics of cosmic
acceleration is just the physics of the cosmological constant.
We have refrained from discussing theories of the
cosmological constant (for reviews, see
Refs.~\cite{Weinberg:1988cp,Carroll:2000fy}), but we make
an exception for the recently developed landscape scenario
\cite{Bousso:2000xa}.

Like quintessence, the landscape scenario allows for a range of
possible values for the vacuum energy.  Unlike
quintessence, these possibilities are arranged
in a ``discretuum,'' rather than a continuum, of values.  The
spacing between these values is comparable to the observed value
of the cosmological constant.  Additional arguments, which we do
not discuss here, can be provided to explain why we live in one
the lowest of these.

To understand the idea, recall that the electromagnetic field $F_{\mu\nu}$ is
a two-form (an antisymmetric rank-two tensor) sourced by a
charge $e$ that follows some worldline.  In 1+1 dimensions (or
equivalently, between two parallel plates), the electric field
and its energy density are constant, and quantization
of the electron charge $e$ implies that both the field and
energy density are quantized, the latter taking on values $\rho
\propto n^2 e^2$, where $n$ is an integer.

Similarly, a four-form field $F_{\mu\nu\rho\sigma}$ in 3+1
spatial dimensions is sourced by coupling to an
``electrically'' charged membrane (a 3-brane), and in string
theory, there are also analogs of magnetic charges (5-branes).
Quantization conditions, analogous to the Dirac quantization
condition in electromagnetism, then require that the field, and
the associated energy density, take on discrete values, $\rho
= (1/2) n^2 q^2 \Mpl^4$
\cite{Abbott:1984qf,Brown:1987dd,Bousso:2000xa}.

Suppose now that there is a ``bare'' cosmological constant
$\lambda =O(\Mpl^2)$ which, for the sake of argument,
may be negative.  Then the effective cosmological
constant $\Lambda$ can take on values $\Lambda=\lambda + 4 \pi
n^2 q^2 \Mpl^2$.  There are thus an infinite range of possible
values of $\Lambda$.  The requirement that there be one that is
$\Lambda \lesssim 10^{-120}\, \Mpl^2$ requires $q\lesssim
10^{-120} \lambda^{1/2} \Mpl^{-1}$; i.e., there is still a fine-tuning
problem. Put another way, if $q\sim 1$, then the closest that
$n^2 q^2$ will get to $-\lambda/\Mpl^2$ is $\sim 1$, or in other
words, the density of states is constant in $n$.

However, in string theory, there may be a large number $J$ of
four-form fields; e.g., a typical value may be $J\simeq100-500$.  If
so, then the cosmological constant takes on values $\Lambda =
\lambda + 4 \pi \sum_i n_i^2 q_i^2 \Mpl^2$.  Taking all $q_i=q$,
for the sake of argument, each combination
$\{n_1,n_2,\cdots,n_J\}$ describes a different vacuum with a
contribution $\lambda_n\equiv 4\pi q^2 \Mpl^2 \sum_i
n_i^2$ to the vacuum-energy density.  The number of states with
$n^2\equiv \sum_i n_i^2$ in the range $n^2 \to n^2+ dn^2$ is
$(dN/dn^2)dn^2$, where $(dN/dn^2)= (2\pi)^{J/2}n^{J-2} [2
\Gamma(J/2)]^{-1}$, the density of states, is proportional to
the area of an $J$-sphere of radius $n$.  The typical
spacing between states is thus $4\pi q^2 \Mpl^2 \Delta(n^2)$
where $\Delta(n^2) = (dN/dn^2)^{-1}$.  If we assume
$\lambda_n\simeq \Mpl^2$, then $n^2\simeq (4\pi q^2)^{-1}$.
Taking $4 \pi q^2 \simeq0.01$, we find $\Delta(n^2) \simeq
10^{-120}$ for  $J\simeq 200$.  Thus, the presence of many
four-form fields allows for far more closely spaced levels in the
cosmological-constant discretuum, and thus explains how a value
$10^{-120}\, \Mpl^2$ may arise in string theory.  

\section{The Observational Way Forward}
\label{sec:observations}

\subsection{The expansion history}

The evidence for dark energy or modified gravity
comes from measurements that probe the expansion history of the
Universe, and extensions of these measurements
provide perhaps the most promising avenues for further empirical
inquiry.  Current data show that the cosmic expansion is
accelerating and constrain the dark-energy density to within a
few percent.  If we assume the equation-of-state parameter $w_Q$
is constant, then it is constrained to be within 12\% of $-1$ (at the
95\% C.L.) \cite{Komatsu:2008hk}.

The question now is whether cosmic acceleration is due
just to a cosmological constant, or whether there is something
more interesting going on.  Thus, a number of avenues are being
pursued to measure $w_Q$ more precisely to see whether it can be
shown to be different from $-1$.  These probes
have recently been reviewed thoroughly by the Dark Energy
Task Force (DETF) \cite{Albrecht:2006um} and in
Refs.~\cite{Frieman:2008sn,Linder:2008pp}, so we simply summarize them here.
In principle, the expansion history can be determined
with a variety of cosmological observations (e.g.,
quasar-lensing statistics, cluster properties,
the Lyman-alpha forest, Alcock-Paczynski test, direct
measurements of the age of the Universe, etc.).  However, the
DETF focused upon supernovae, galaxy-cluster
abundances, baryon acoustic oscillations, and weak gravitational
lensing, reflecting a rough consensus in the community that
these four currently provide the most promising avenues.  We
caution, however, that there may still be room for new ideas.
Either way, it is generally agreed that given systematic errors
inherent in any particular technique, several complementary
methods will be required to provide cross-checks.

\subsubsection{Supernovae}

Supernovae have played a crucial role in establishing cosmic
acceleration, and they are likely to provide even more precise
constraints on the expansion history in the future.  
The
supernovae used here are Type Ia supernovae, explosions
powered by the thermonuclear detonation of a white dwarf when its
mass exceeds the Chandrasekhar limit.  These explosions can be
distinguished from those produced by other mechanisms (e.g.,
Type II supernovae, powered by iron-core collapse in
supergiants) from the details of their spectra and light
curves.  The fact that the star ignites very rapidly after
exceeding the Chandrasekhar limit implies that Type Ia
supernovae should be good standard candles.  Thus,
their observed brightness provides the luminosity distance
$d_L(z)$.  Measurements
support this simple notion, and details of the spectra and light
curves can be used to correct for relatively small changes in the
supernova luminosities.

Supernova searches will be particularly valuable if they can reach
redshifts $z\sim1$, where the effects of different $w_Q$ values
become most dramatic (see Fig.~\ref{figure:expansion}.  Progress
with supernovae will require
improved systematic-error reduction, better theoretical
understanding of supernovae and evolution effects, and greater
statistics.  Both ground-based and space-based
supernova searches can be used to determine the expansion
history.  However, for redshifts $z\gtrsim1$, the principal optical
supernova emission (as well as the characteristic silicon
absorption feature) gets shifted to the infrared which is
obscured by the atmosphere, and this provides (much of) the case
for a space-based observatory.

\subsubsection{Baryon acoustic oscillations}

In recent years, baryon acoustic oscillations (BAO) have become
increasingly attractive as a possibility for determining the
expansion history.  The acoustic oscillations seen in the CMB
power spectrum are due to oscillations in the photon-baryon
fluid at the surface of last scatter.  The dark matter is
decoupled and does not participate in these oscillations.
However, since baryons contribute a non-negligible fraction of
the nonrelativistic-matter density, oscillations in the
baryon-photon fluid get imprinted as small oscillations in the
matter power spectrum at late times
\cite{Eisenstein:1997gf,Seo:2003pu}.
These oscillations have now been detected in galaxy
surveys \cite{Eisenstein:2005su}.  The
physical wavenumber at which these oscillations occur is
well understood from linear perturbation theory, and so they
provide a standard ruler.  Thus, baryon acoustic oscillations
measure the angular-diameter distance $d_A(z) =(1+z)^{-2}
d_L(z)$.  Measurement of clustering along the line of sight may
also provide information on the expansion history $H(z)$.
Issues with BAO include
nonlinear evolution of the acoustic peaks in the matter power
spectrum and systematic and astrophysical effects
\cite{Pritchard:2006ng} that could mimic features in the power spectrum.

\subsubsection{Cluster abundances}

Galaxy clusters are the largest gravitationally bound objects in
the Universe.  The spatial density of clusters in the Universe
can be determined from models of structure formation.  The
observed number of clusters depends on the spatial density as
well as on the volume per unit solid angle on the sky and per
unit redshift interval  \cite{Haiman:2000bw}.  This volume
depends on the quantity $[H(z)(1+z)]^{-1}$, and so clusters
measure the expansion history $H(z)$.   

The theories predict the cluster
abundance as a function of the cluster mass.  The trick, then,
is to obtain the cluster mass from the cluster
observables---namely, the luminosity and temperature of the
x-ray emission, the Sunyaev-Zeldovich effect
\cite{Carlstrom:2002na}, cluster dynamics,
and/or the effects of weak
gravitational lensing by the cluster on background galaxies.
There is now a large industry that amalgamates theory,
simulations, and multiwavelength cluster
observations in an effort to develop a reliable cluster-mass
indicator.  

\subsubsection{Weak lensing}

Weak gravitational lensing by large-scale density fluctuations
along the line of sight to distant galaxies can distort the
images of those galaxies \cite{Refregier:2003ct}. Large-distance
correlations in the
mass thereby induce long-distance correlations in the observed
ellipticities of the distant galaxies.  
Measurements of these ellipticity correlations can thus be
used to determine the power spectrum of the mass as a function
of angular wavenumber.  If the  power spectrum is already known,
e.g., from the CMB, as a function of the physical wavenumber,
then the observed amplitude determines the physical wavenumber
corresponding to an angular wavenumber.  Thus, weak lensing
measures the angular-diameter distance $d_A(z)$.
Weak lensing probes the gravitational potential, and thus the total mass,
unlike galaxy surveys, which use luminous galaxies to trace the
mass distribution.  The challenge
with weak lensing is to understand the subtle experimental
effects that might mimic weak-lensing--induced ellipticity correlations.
There may also be intrinsic alignments of the galaxies
\cite{Catelan:2000vm} that might look like a weak-lensing signal.

\subsubsection{Other probes of the expansion history}

There may be other ways to measure the expansion history.  If
the ages of stellar populations can be obtained from their
spectra at a variety of redshifts, then the expansion rate
$dz/dt$ may be obtained directly \cite{Jimenez:2001gg}.  There
may be other luminous standard candles; for example,
the gravitational-wave signal from supermassive-black-hole binaries
\cite{Holz:2005df} may give a new method to determine luminosity distance if a
suitable measure of redshift can be obtained from an optical counterpart.
It has also been recently suggested that by
comparing the biases and redshift-space distortions for two
different galaxy populations, constraints to $D(z)$ and $H(z)$
may be obtained \cite{McDonald:2008sh} in a way that is limited
ultimately by the number of galaxies, rather than the number of
Fourier modes in the density field.

\subsection{Growth of structure}

The rate of growth of density inhomogeneities [i.e., the linear-theory
growth factor $D(z)$] depends on the cosmic expansion rate.
Moreover, different theories that predict the same background cosmic
evolution may lead to different rates of perturbation
growth.  For example, DGP theories are expected to have a
significant effect on $D(z)$, and we have discussed above a
phenomenological approach (parametrized-post-Friedmann) to the
growth of perturbations in alternative-gravity theories.
Of the four avenues discussed above, clusters, BAO, and weak lensing may
also provide measurements of $D(z)$, in addition to measurements
of $H(z)$.

\subsection{Lorentz violation and other tests}

The new physics, gravitational or otherwise, implied by cosmic
acceleration may have other observable/experimental
consequences, apart simply from its effect on cosmic expansion.
For example, we have discussed Solar System tests of
alternative-gravity theories for cosmic acceleration and the
differing effects of various models on the growth of large-scale
structure.

Tests of Lorentz violation provide another avenue.  The rest
frame of the CMB provides us with a preferred frame in the
Universe.  Since a cosmological constant has the same density in
every inertial frame, it can manifest no effects of Lorentz violation.
If, however, $w\neq-1$, either due to dark energy or modified
gravity, and if that new physics is somehow coupled
non-gravitationally to ordinary matter, then the preferred
cosmological frame may show up in tests of Lorentz violation.
Typically, however, we expect these violations to be extremely
small by laboratory standards.  First of all, dark-energy fields
must be exceedingly weakly coupled to Standard Model particles
if they are to remain dark.  Moreover, the timescale for
evolution of these fields is the Hubble time, far longer than
laboratory timescales.

Cosmological observations might allow for the
experimental timescale to be comparable to the Hubble time.
For example, Ref.~\cite{Carroll:1998zi} pointed out that if quintessence
couples to the pseudoscalar of electromagnetism, there
will be a uniform rotation of the linear polarization of
photons propagating over cosmological distances.  This could be
probed by looking for a mean misalignment between the linear
polarization of cosmological radio sources with the position
angles of their images.  It can also be tested by looking for
the parity-violating polarization correlations it produces in the CMB
polarization \cite{Lue:1998mq}.  

In addition to these and laboratory tests of Lorentz violation,
preferred-frame effects in gravitational physics may also arise
if the quintessence field couples in some nontrivial way
\cite{Graesser:2005bg}.  Eotvos-like experiments may also be
used to search for couplings of ordinary matter to the
quintessence field.  If cosmic
acceleration is due to a scalar-tensor theory, then the variable
Newton's constant implied by the theory may suggest that other
fundamental constants vary with time \cite{Uzan:2002vq}.

It is easy to speculate how various dark-energy theories may give
rise to Lorentz violation, preferred-frame effects, or variation
of fundamental constants.  But in the absence of any clear
front-runner theories, it is much more difficult to say which, if
any of these, will be more constraining.

\section{Conclusions}

Cosmic acceleration provides an intriguing puzzle. Occam's razor
suggests that the phenomenon may be explained simply by a
cosmological constant.  This may be an acceptable
phenomenological explanation, but it would be more satisfying to have a
physical explanation for the observed value of $\Lambda$. The
unexpectedly small
value inferred for $\Lambda$ leads us to suspect that instead the apparent
cosmological constant may be the false-vacuum energy associated with the
displacement of some field from its minimum and/or that there may be new
gravitational physics beyond Einstein's general relativity. Plenty of interesting
ideas for dark energy and alternative gravity have been conjectured, but there is no
clear front runner. The models are all toys, awaiting any new, corroborating or
contraindicating evidence.

Some have argued that no new physics is required, that
nonlinear behavior in general relativity may exhibit subtleties
that allow for an accelerated expansion.
For example, Ref.~\cite{Kolb:2005me} proposed that 
superhorizon perturbations may induce
accelerated expansion in our observable Hubble patch.
This idea has been disproved
\cite{Hirata:2005ei,Geshnizjani:2005ce,Flanagan:2005dk},
but it has not yet been disproven that sub-horizon
nonlinearities may explain the observations. Alternatively, it
has been suggested that the luminosity-distance--redshift data
can be explained if we reside at the center of a
Gigaparsec-scale void in an otherwise Einstein-de Sitter Universe.
But such a radially-inhomogeneous, anti-Copernican scenario conflicts with other
observations \cite{Goodman:1995dt,Caldwell:2007yu}. 
Any future proposals that attempt to dispense with new physics
will have to explain the vast catalog of
phenomena already explained by the standard cosmological model.

New theories of gravitation can in principle work.  However, it
has proved to be more difficult than may have originally been
anticipated to alter gravity to explain cosmic acceleration
without violating Solar System constraints.  The scalar-tensor
or $f(R)$ theories that do succeed seem contrived and/or
manifest themselves in a way that is virtually indistinguishable
from a cosmological constant.  Braneworld scenarios introduce
the possibility of interesting gravitational physics in the
Solar System and in large-scale structure, but the simplest
models must be ornamented with additional ingredients to work.
Generally, alternative-gravity theories that alter the
long-range $1/r^2$ force law may have profound implications for
a variety of astrophysical systems, few of which have yet been
explored carefully.

The simplest paradigm, quintessence,
does not suffer from instabilities, and can be viewed as an
effective theory for more complicated models.  Quintessence
models do require small parameters and/or finely-tuned initial
conditions, and they do not address the coincidence problem.
Still, the resemblance of some quintessence fields to both
fundamental or composite scalars appearing in existing models of
physics beyond the Standard Model give the hope that new
particle discoveries, at the Large Hadron Collider or beyond, may provide the
clues to connect this dark energy field to the world of luminous
matter.

The next step for cosmological studies should be to determine if
$w_Q$ departs significantly from $-1$.  If it does, then the
step beyond that will be to measure its time evolution $w_a$.  The
$w_0$-$w_a$ measurement may then tells us something qualitative
about dark-energy dynamics (e.g., thawing or freezing
potentials).  If so, we can then go further from there.

\bigskip

\section*{Acknowledgments}
We thank S.~Carroll, A.~Erickcek, J.~Frieman, T.~Smith, and A.~Weinstein for
useful comments on an earlier draft.  This work was supported at
Caltech by DoE DE-FG03-92-ER40701 and the Gordon and Betty Moore
Foundation and at Dartmouth by NSF AST-0349213.

\section*{Appendix A: Acronyms}

\renewcommand{\labelitemi}{}

\begin{itemize}
\item BAO: Baryon Acoustic Oscillations
\item CMB: Cosmic Microwave Background
\item DETF: Dark Energy Task Force
\item DGP: Dvali-Gabadadze-Porrati
\item GR: General Relativity
\item JBD: Jordan-Brans-Dicke
\item MaVaN: Mass Varying Neutrino
\item PNGB: Pseudo-Nambu Goldstone Boson
\item PPN: Parametrized Post-Newtonian
\end{itemize}

\section*{Appendix B: Key Terms}

\begin{itemize}
\item {\it Dark Energy}: A negative-pressure fluid comprising
$\sim75\%$ of the cosmic energy budget, postulated to account
for the accelerated cosmic expansion.

\item {\it Quintessence}: A dynamical dark energy. Literally, the
fifth element in the cosmic energy budget, beyond radiation,
baryons, neutrinos, and dark matter.

\item {\it Equation-of-State Parameter}: The ratio of the homogeneous
pressure to the energy density, and denoted $w = p/\rho$.

\item {\it Braneworld}: Scenario in which Standard Model fields are
confined to a membrane in a higher-dimensional spacetime, but
gravity propagates everywhere.

\item {\it Landscape scenario}: The idea that string theory predicts a
huge number of false vacua with different but closely spaced
vacuum-energy densities.
\end{itemize}

\section*{Appendix C: Highlighted References}

\begin{itemize}

\item Refs.~\cite{Perlmutter:1998np} \& \cite{Riess:1998cb}:
These two papers report the first direct evidence,
from supernova measurements of the luminosity-distance--redshift
relation, for an accelerated cosmic expansion.

\item Refs.~\cite{Albrecht:2006um} \& \cite{Frieman:2008sn}:
These two articles provide the most up-to-date and
detailed reviews of observational probes of the expansion
history.

\item Ref.~\cite{Copeland:2006wr}:    This is a comprehensive
and detailed review of dynamical models of dark energy.

\item Ref.~\cite{Carroll:1998zi}:      This article provides a
cogent explanation of the difficulties in building a dark cosmic
scalar field in a realistic model of particle physics.

\item Ref.~\cite{Lue:2005ya}:     This article provides a clear
and detailed recent review of DGP gravity and cosmology.

\item Ref.~\cite{Weinberg:1988cp}:   This is a classic review on
the cosmological constant.

\end{itemize}

\section*{Appendix D: Summary Points}

\begin{itemize}

\item 1. The cosmic expansion is observed to be accelerating.

\item 2. The physical mechanism responsible for the cosmic
acceleration is unknown. Interpreting the observational and
experimental evidence in the context of Einstein's general
relativity, the causative agent appears to be an exotic fluid,
referred to as ``dark energy,'' with negative pressure.

\item 3. A cosmological constant is equivalent to such a fluid with a
constant energy density.  However, the value of this energy
density, in units where $G=c=\hbar=1$ is $10^{-120}$, and there
is no good explanation for the smallness of this value.

\item 4.  Quintessence postulates that the dark energy
is due to the negative pressure associated with the displacement
of some new scalar field from the minimum of its potential. Such
theories generally predict an equation-of-state parameter for
dark energy of $w_Q\neq-1$, as opposed to the cosmological
constant, which has $w_Q=-1$.

\item 5.  Other explanations for cosmic acceleration propose that a
new gravitational theory supplants Einstein's general relativity
on cosmological scales.  However, new theories are tightly
constrained by precision tests of gravitation within the Solar
System.

\item 6. In the absence of a clear front-runner theory, most efforts
are directed towards refining measurements of the cosmic
expansion history to determine more precisely the value of $w_Q$.

\item 7. A combination of cosmological observations are expected to
gain the most traction towards understanding the physics of
cosmic acceleration. Most attention has been focused on four
techniques: supernovae, baryon acoustic oscillations, cluster
abundances, and weak lensing.
\end{itemize}

\section*{Appendix E: Problems with the Cosmological Constant}

A cosmological constant with $\Lambda=3 \Omega_\Lambda
H_0^2/c^2$ provides a phenomenological description of dark
energy; it implies that the vacuum ``weighs'' something; i.e.,
that the vacuum gravitates.  However, there is no physical
understanding for why empty space would act as a source for the
gravitational field. The particle-physics vacuum contributes
an effective cosmological constant, but with an energy density
many orders of magnitude larger than is observed. This gross
mismatch between theory and observation---noted both by Pauli
\cite{Straumann:2008aa} and Zeldovich
\cite{Zel'dovich:1968zz}---is one of the deepest physical
enigmas of our time.  In quantum field theory, renormalization
allows us to reset the energy density of the vacuum to zero, and
for many years it was generally assumed that some mechanism made
this cancellation precise and stable.  However, the discovery of
cosmic acceleration suggests that the cosmological constant
is (in the absence of quintessence of some
alternative-gravity explanation for cosmic acceleration) small,
but non-zero, and this has now changed the character of the
cosmological-constant problem.  If the observational trend
continues to favor dark energy with $w_Q$ consistent with $-1$,
the challenge will be to explain why the cosmological constant
is so small, yet non-zero.

\section*{Appendix F: Extra resources}

\begin{itemize}

\item J.~P.~Uzan, ``The acceleration of the universe and the
     physics behind it,'' Gen.\ Rel.\ Grav.\  {\bf 39}, 307
     (2007) [arXiv:astro-ph/0605313].

\item R.~Durrer and R.~Maartens,
     arXiv:0811.4132 [astro-ph].

\item R.~R.~Caldwell,
     Phys.\ World {\bf 17}, 37 (2004).
     http://physicsworld.com/cws/article/print/19419

\item S.~Nobbenhuis,
     ``Categorizing Different Approaches to the Cosmological Constant Problem,''
      Found.\ Phys.\  {\bf 36}, 613 (2006)
       [arXiv:gr-qc/0411093].

\item http://universe.nasa.gov/science/darkenergy.html

\end{itemize}

\section*{Appendix G: Future Issues}

\begin{itemize}

\item Will future results from the Large Hadron Collider have any
impact on dark-energy theory?  Could the discovery of
supersymmetry, a nonstandard Higgs, or large extra dimensions
change the way we think about dark energy?

\item Will string theory make a robust prediction for the cosmological
constant? or perhaps otherwise explain the physics of cosmic
acceleration?

\item Can an elegant and consistent modification to general
relativity be found to explain cosmic acceleration while still
satisfying Solar System constraints?

\item Will there be NASA and ESA satellite missions to study dark
energy within 5-10 years?

\item How much will ground-based observations and experiments refine
our knowledge of the physics of cosmic acceleration?

\item Will new connections between other probes of new physics (e.g.,
dark-matter searches, gravitational waves, probes of gravity on
submillimeter scales? Lorentz-invariance violation) and dark
energy be found?

\item The relevant future observations include measurements of the
cosmic expansion history with greater accuracy and studies of
the growth of large-scale structure.  More work must be done to
determine the best avenue forward will be in this regard.
\end{itemize}


\end{document}